\documentclass[]{interact}
\usepackage{lmodern}
\usepackage[caption=false]{subfig}
\usepackage[colorlinks=true,citecolor=blue,urlcolor=blue,linkcolor=blue]{hyperref}

\usepackage[numbers,sort&compress]{natbib}
\bibpunct[, ]{[}{]}{,}{n}{,}{,}
\makeatletter
\def\NAT@def@citea{\def\@citea{\NAT@separator}}
\makeatother

\theoremstyle{plain}

\theoremstyle{definition}

\theoremstyle{remark}

\newcommand{\kv}{{\bf k}}
\newcommand{\qv}{{\bf q}}
\newcommand{\down}{\downarrow}
\newcommand{\up}{\uparrow}
\begin{document}


\title{Gaps in unconventional superconductors}

\author{
\name{Andreas Kreisel
}\affil{Department of Physics and Astronomy, Uppsala University, Box 524, 751 20 Uppsala, Sweden}
}

\maketitle

\begin{abstract}
The energy gap is one of the defining properties of a superconductor and appears because the quasiparticle excitations radically change once a metallic system enters the superconducting state. Unconventional superconductors typically exhibit a non-uniform gapping and therefore show physical effects that are subject to intense research. In this review, we provide an overview of concepts needed to understand how unconventional superconductors are different from conventional ones, where the gapping comes from and how it expresses itself in experimentally accessible quantities. This should give the basis to understand current open research questions and navigate the recent literature that tend to contain contradictory conclusions.
\end{abstract}

\begin{keywords}
Superconductivity; energy gap; Cooper pair; BCS theory; spin fluctuations; phase-sensitive probes
\end{keywords}

\section{Introduction}

Superconductivity is a phase of matter in which the wavefunction is phase coherent over macroscopic scales, so quantum mechanical effects are not confined to atomic length scales. This makes basic research on and applications of superconductors a fascinating topic.
Still, many puzzles connected to unconventional superconductivity are unsolved. In conventional superconductors, electron pairing is mediated by lattice vibrations (phonons) and is well described by the BCS theory as first laid out by Bardeen, Schrieffer and Cooper\cite{BCS1957}. In contrast, pairing in unconventional superconductors was discovered in heavy-fermion materials\cite{Steglich1979} and is believed to arise from alternative pairing mechanisms, often originating from electron-electron interactions.

The term unconventional superconductivity is also used in a different sense to describe superconducting states with a nontrivial order parameter symmetry. In this context, a superconductor is considered unconventional (more precisely nontrivial) when its order parameter has a lower symmetry than the underlying crystal lattice, leading to anisotropic or sign-changing pairing states. Although these two uses of the term often coincide, they are conceptually distinct, and we will distinguish these in the following.

As a starting point, we want to describe the phenomenology of a superconductor that applies to conventional superconductors as well as to unconventional superconductors.
\begin{enumerate}
 \item Superconductivity was discovered more than 100 years ago when cooling down the metal Hg to low temperatures. Below the critical temperature ($T_c$), the resistivity of a superconductor is truly zero, allowing the current to flow without energy loss. Electric currents can circulate indefinitely in a superconducting loop without any applied voltage. These currents have been experimentally observed to persist for long times.
\item The Meissner effect \cite{Meissner1933}, i.e. that superconductors exhibit perfect diamagnetism, is another phenomenology that is inherently connected to superconductivity and distinguishes it from other phases: Superconductors expel magnetic fields from their interior by generating screening currents near the surface. This screening occurs on a characteristic length scale $\lambda$ where the field typically decays as $B=B_0 e^{-r/\lambda}$ with distance $r$ from the surface.
\item  Superconductors open an energy gap in the single particle excitation spectrum around the Fermi surface.
However, having an energy gap alone does not guarantee superconductivity, and some superconductors, especially many unconventional superconductors with nontrivial order parameter, do not have a complete gap.
\end{enumerate}

The superconducting state is a distinct phase which can be described by an order parameter $\Delta_{\mathbf{k}}$. It is generically zero above the ordering temperature $T_c$ and acquires a nonzero value in the superconducting state. For unconventional superconductivity it is important to note that $\Delta_{\mathbf{k}}$ can depend on the crystal momentum $\kv$ and other quantum numbers $\alpha$. This dependence determines the symmetry of the superconducting gap and provides a natural basis for classifying different pairing states.
To set the stage for the discussion of the superconducting order parameter, we introduce some important concepts of solid state physics, the theoretical description of electrons as fermions and how to calculate expectation values of fermionic systems at finite temperature. These concepts will be used to explain the properties and formation of the superconducting order parameter. Further, we discuss symmetry-based classifications of superconducting states on the example of a square lattice system that has relevance for a number of real materials. Focusing on spin-fluctuation-mediated superconductivity, we will review the basic ideas to theoretically understand the emergence of unconventional superconductivity.
To further connect to recent research questions, we give an overview of relevant experimental probes used to identify unconventional pairing and explain how the underlying gap symmetry and pairing structure are reflected in measurable quantities.

\subsection{Superconducting order parameter}
The ground state of a superconductor is characterized by the anomalous expectation value 
$\langle c_{-\mathbf k,\downarrow}c_{\mathbf k,\uparrow}\rangle$
as first introduced in the theory by Bardeen-Cooper-Schrieffer (BCS)\cite{BCS1957}. If this quantity vanishes, the system is said to be in the normal (non-superconductive) state, if it is nonzero, the system is in the superconducting state. 
If restricting to pairing of electrons with opposite momenta, so-called zero momentum Cooper pairs, one can consider the quantity
\begin{equation}
  b_{\mathbf{k} \alpha\beta}=\langle c_{-\mathbf k,\alpha}c_{\mathbf k,\beta}\rangle\label{eq_pair_corr}
\end{equation}
where $\mathbf k$ is the crystal momentum and $\alpha$ and $\beta$ label other quantum numbers of the electrons described by operators $c^\dagger_{\mathbf k,\beta}$ and $c_{\mathbf k,\beta}$ creating and annihilating an electron at momentum $\mathbf k$ with quantum number $\beta$.
For the reader not familiar with the description of electronic states, we refer to Sec.~\ref{sec_app} to the concept of crystal momentum, the properties of fermions, and the definition of the expectation value. Considering quantum states with a fixed number of electrons, it becomes clear that removing the electron with $\mathbf k,\beta$ and the electron with $-\mathbf k,\alpha$, i.e. removing a Cooper pair, will lead to a state with two fewer electrons. This state is orthogonal to the original state and therefore the expectation value vanishes for a normal metal. From this perspective, the superconducting ground state is quite different, namely it consists of superpositions of states with different electron numbers. Only in this case, the expectation value can become nonzero. 

The first additional observation is that this is the expectation value of a non-Hermitian operator because $(c_{-\mathbf k,\alpha}c_{\mathbf k,\beta})^\dagger=c^\dagger_{\mathbf k,\beta}c^\dagger_{-\mathbf k,\alpha}\neq c_{-\mathbf k,\alpha}c_{\mathbf k,\beta} $, and therefore not a physically measurable quantity. The second observation is that the Pauli exclusion principle, i.e. $ c_{-\mathbf k,\alpha}c_{\mathbf k,\beta}=- c_{\mathbf k,\beta}c_{-\mathbf k,\alpha}$ as explained in Sec.~\ref{pauli},  imposes constraints to make the total pair wavefunction antisymmetric under particle exchange, i.e.
$b_{\mathbf {k}\alpha\beta}=-\langle  c_{\mathbf k,\beta} c_{-\mathbf k,\alpha}\rangle=-b_{-\mathbf {k} \beta\alpha}$. As a consequence, the symmetry properties of the expectation value are restricted, linking its momentum to the other quantum numbers. For a more detailed discussion of this point, see Ref.~\cite{Sigrist_notes}.
In this review, we will only discuss the case without additional quantum numbers, i.e. the single band case. 
There are two possibilities to get the overall minus sign: Either the momentum function is even, i.e. invariant under $\kv \rightarrow -\kv$ and the exchange of the spin quantum numbers needs to give the minus sign, \emph{singlet pairing} or the momentum function is odd and picks up the minus sign, then the exchange of the spin quantum numbers needs to be even, \emph{triplet pairing}.

\subsection{Mean field description of superconductivity}
\begin{figure}[tb]
\centering
\includegraphics[width=0.45\linewidth]{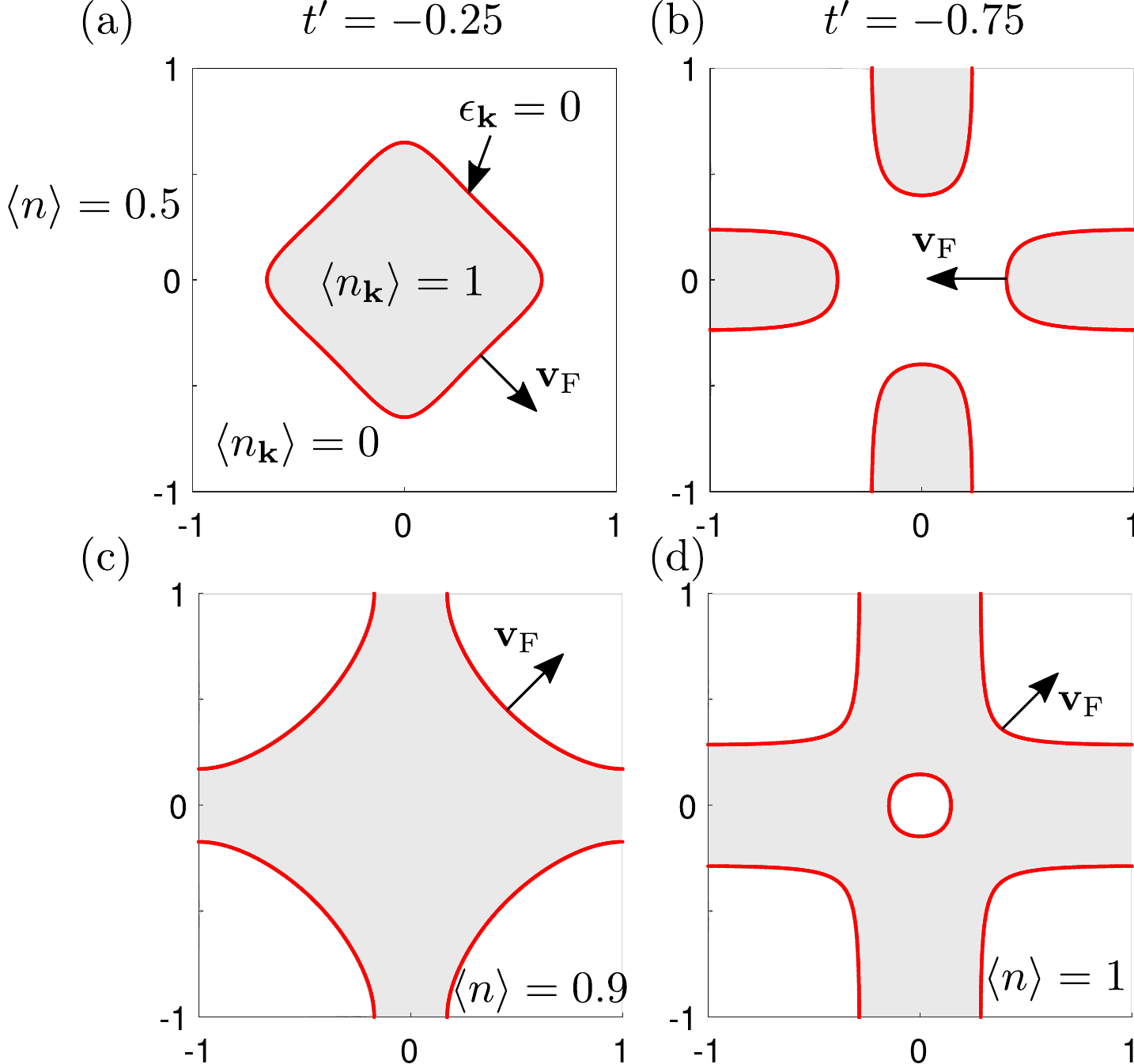}
\caption{The single band model can give four different Fermi surface topologies (red lines) when choosing the parameter $t'$ (columns) and the filling $\langle n\rangle$ accordingly (a-d); filled states in the first Brillouin zone with momenta $k_i \in (-\pi/a \ldots \pi/a]$ are indicated as light gray area and the direction of the Fermi velocity $\mathbf{v}_\mathrm{F}$ is indicated with arrows.}
       \label{fig_lifshitz}
\end{figure}

After these general symmetry considerations, we want to describe how the anomalous expectation value can become nonzero and what this means for the quantum states and the excitation gap.
As a model system, we will use a single band tight binding model with dispersion
\begin{align}
 \epsilon_\kv=-2t(\cos k_x +\cos k_y)-4t'\cos k_x \cos k_y -\mu, \label{eq_disp_single}
\end{align}
where $t$ and $t'$ are hopping amplitudes and $\mu$ is the chemical potential; see Sec.~\ref{sec_single_band} for more details of this model. In Fig.~\ref{fig_lifshitz} (a), we present the 4 possible Fermi surface topologies to demonstrate the flexibility of this dispersion to model various interesting electronic band structures, although a large value of $|t'|$ cannot be motivated for real materials.

\begin{figure}[tb]
\centering
\includegraphics[width=\linewidth]{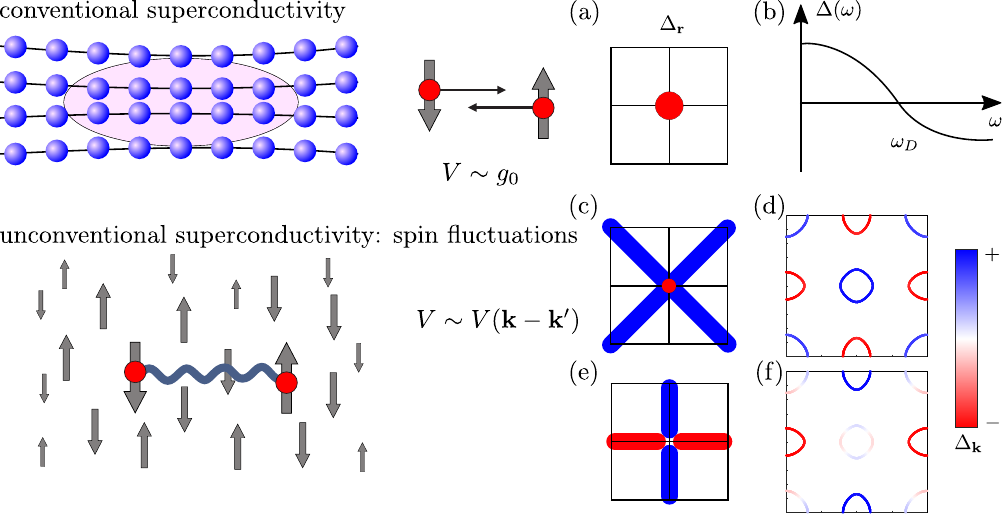}
\caption{{\bf Conventional pairing:} The effective attractive interaction (black arrows) between two electrons (red dots) arises when the movement of the first deforms the lattice of positively charged atomic cores (light red region). The second electron is attracted to an effective positive charge accumulation.
The order parameter $\Delta_{\mathbf r}$ is just onsite ${\mathbf r}=0$ (a), but with a sign change at the characteristic frequency $\omega_D$ \cite{Scalapino2012,Coleman2015}.
{\bf Unconventional pairing with the example of spin fluctuations:} The effective interaction arises from a single electron (with spin) polarizing the cloud of conduction electrons by the Coulomb interaction. If the polarization is antiferromagnetic, i.e. correlations of neighboring electrons on the lattice tend to have the opposite spin direction, the second electron can lower its energy in the polarized region of the first electron, leading to an effective interaction (wiggly line)\cite{ChubukovHirschfeld2015,Scalapino2012,Maiti2013,Norman2011}. The pairing interaction is then momentum dependent and induces a momentum dependent order parameter $\Delta_\kv$. The complication of unconventional pairing mechanisms is that multiple order parameters can be realized; here two examples are shown:
The $s_\pm$ state has the full symmetry of the lattice (c), but
exhibits pairing to the next nearest neighbors. A sign change
of the order parameter $\Delta_{\mathbf k}$ on the Fermi surface occurs (d).
The $d_{x^2-y^2}$ order parameter is a \emph{nontrivial} pairing state since it has lower symmetry than the lattice. Under a 90 degree rotation, it changes sign since  the horizontal bond orders have opposite sign than the vertical ones (e). The same symmetry property under a 90 degree rotation is given in momentum space, such that $\Delta_\kv$ is forced to vanish at the nodal lines. The order parameter has symmetry-enforced nodes on the Fermi surface and very small gap (light colors) on the $\Gamma$ and M pockets (f)\cite{HKM_ROPP}.}
       \label{fig_el_phonon_SF}
\end{figure}

The theoretical description of superconducting pairing in BCS theory starts from a metallic normal state with a Fermi sea, i.e. filled states below zero energy and an assumed attractive interaction of strength ($g_0$) acting within an energy window ($\omega_D$) around the Fermi level. This attraction leads to an instability of the Fermi sea toward a superconducting state.
Usually, this effective model is motivated by the presence of an electron-phonon interaction as mechanism for conventional superconductivity. In detail, the electrons in the Cooper pair overcome the repulsive Coulomb interaction by avoiding each other in time, i.e. the interaction is retarded, see Fig.~\ref{fig_el_phonon_SF}. When considering pair correlations in time, this means that there is a frequency dependence of the order parameter and it changes sign as a function of frequency\cite{Scalapino2012,Coleman2015}.

In the following, we will discuss some essences of unconventional pairing mechanism, where the electrons avoid each other in space rather than in time. In this case, the order parameter acquires a momentum dependence, thus can be nontrivial in view of the symmetry classification of the previous section. When performing a Fourier transformation, the pairing then corresponds to an order parameter that also lives on the bonds (see Fig.~\ref{fig_el_phonon_SF}(c-f)), similar to how the electron hopping yields a momentum dependence in the dispersion when rewriting into the Bloch Hamiltonian as discussed in Sec.~\ref{electronic_states}.

To be concrete, we start from a generic single band Hamiltonian including some (effective) interactions of the form $V(\mathbf k,\mathbf k')$
\begin{align}
 H&=\sum_{\kv,\sigma}\epsilon_{\kv} c_{\kv \sigma}^\dagger c_{\kv\sigma}+\frac{1}{2}\sum_{\kv,\kv'}[
 V(\kv,\kv')
 c_{\kv'\up}^\dagger c_{-\kv'\down}^\dagger
 c_{-\kv\down} c_{\kv\up} 
 + \mbox{H.c.}].
 \label{eq:BCS}
 \end{align} 
H.c. means the Hermitian conjugate, that is constructed by using the usual rules of operator algebra, $(AB)^\dagger=B^\dagger A^\dagger$ and we have extended the band with dispersion $\epsilon_{\kv}$  to include the spin degree of freedom $\sigma=\{\uparrow,\downarrow\}$.

\subsubsection{Mean field decoupling}
The Hamiltonian in Eq.~(\ref{eq:BCS}) contains a four-fermion term that often cannot be solved exactly. However, a very successful approach is to map it onto a formally non-interacting problem, by the mean field approximation. The general strategy is the following: We start from an interacting Hamiltonian
\begin{equation}
 H=H_0+AB
\end{equation}
containing the product of two operators $A$ and $B$ which are for example bilinear in fermion operators and a non-interacting term $H_0$. Next, we rewrite the operators as sum of mean value and deviation from it, $A=\langle A\rangle +\delta A$, $B=\langle B\rangle +\delta B$, insert this in the Hamiltonian, multiply out the square and ignore the term $\delta A\delta B$ of second order in the deviations from the mean. Introducing again the original operators by $\delta A=A-\langle A\rangle $, we obtain
\begin{equation}
 H\rightarrow H_{\mathrm{MF}}=H_0+\langle A\rangle B + A\langle B\rangle -\langle A\rangle\langle B\rangle\,.
 \end{equation}
This Hamiltonian is now non-interacting, as $\langle A\rangle$ and $\langle B\rangle$ are just numbers. However, it depends on the mean values of the two operators as parameters such that one needs to find a self-consistent solution in which the mean values $\langle A\rangle$ and $\langle B\rangle$ are calculated using the eigenvalues and eigenstates of $H_{\mathrm{MF}}$.

\subsubsection{Decoupling in the superconducting channel}
The important insight is now to choose $A=c_{\kv'\up}^\dagger c_{-\kv'\down}^\dagger$ and $B=c_{-\kv\down} c_{\kv\up}$ from the mean field decoupling in anticipation of a finite value of the anomalous expectation value. The remaining steps are some algebra for the Hamiltonian
\begin{align}
 H_{\mathrm{MF}}&=\sum_{\kv,\sigma}\epsilon_{\kv} c_{\kv\sigma}^\dagger c_{\kv\sigma}
+
 \frac{1}{2}\sum_{\kv,\kv'}V(\kv,\kv') \Bigl[
 \langle c_{\kv'\up}^\dagger c_{-\kv'\down}^\dagger\rangle
 c_{-\kv\down} c_{\kv\up}
+ c_{\kv'\up}^\dagger c_{-\kv'\down}^\dagger
 \langle c_{-\kv\down} c_{\kv\up}\rangle
 \notag\\ &\qquad\qquad\qquad\qquad\qquad\qquad\qquad
- \langle c_{\kv'\up}^\dagger c_{-\kv'\down}^\dagger\rangle
 \langle c_{-\kv\down} c_{\kv\up}\rangle
+ \mbox{H.c.} \Bigr],\label{eq_mf}
\end{align}
where the (thermal) expectation values $\langle A\rangle=\langle c_{\kv'\up}^\dagger c_{-\kv'\down}^\dagger\rangle$ and  $\langle B\rangle=\langle c_{-\kv\down} c_{\kv\up}\rangle$ are neither singlet nor triplet pair correlations. In presence of inversion symmetry, it is convenient to introduce the order parameters in the singlet (s) and triplet (t) channel,
\begin{align}
 \Delta_{\kv}^{s/t}=- \sum_{\kv'}V^{s/t}(\kv,\kv')\langle c_{-\kv'\down} c_{\kv'\up}\rangle\label{eq_gap}\,,
\end{align}
where the expectation value has to be calculated using the eigenstates of Eq.~(\ref{eq_mf}) and we need to examine under which circumstances this order parameter can become nonzero to enter the superconducting state.
As a side remark, we introduce symmetric (singlet) and antisymmetric (triplet) part of the interaction via
\begin{equation}
  V^{s/t}(\kv,\kv') =\frac{1}{2}[V(\kv,\kv')\pm V(-\kv,\kv')]. \label{eq_pairing_symm}
\end{equation}
After this algebra, we obtain a Hamiltonian that contains terms with two fermionic creation operators (or two annihilation operators),
\begin{align}
 H_{\mathrm{MF}}=\sum_{\kv,\sigma}\epsilon_{\kv} c_{\kv\sigma}^\dagger c_{\kv\sigma}-\sum_\kv \Delta_{\kv}^{s/t*}c_{-\kv\down} c_{\kv\up}\mp\sum_\kv \Delta_{\kv}^{s/t}c_{\kv\up}^\dagger c_{-\kv\down}^\dagger+\text{const.}\label{eq_bcs_mf}
\end{align}
The Pauli principle constrains the singlet order parameter to be even in the momentum $\kv$, $\Delta^s_\kv=\Delta^s_{-\kv}$ (even parity) and the triplet order parameter to be odd in the momentum $\kv$, $\Delta^t_\kv=-\Delta^t_{-\kv}$ (odd parity).
On this Hamiltonian, we can see that an eigenstate cannot be of a fixed number of electrons because the term with two creation operators would generate a state with two additional electrons.  The order parameter term is also not invariant under a $U(1)$ transformation $c_{\kv\sigma} \rightarrow e^{i\phi_0}c_{\kv\sigma}$, $c^\dagger_{\kv\sigma} \rightarrow e^{-i\phi_0}c^\dagger_{\kv\sigma}$  with a global phase $\phi_0$ because two annihilation operators will acquire the phase $e^{2i\phi_0}$. The (complex-valued) order parameter chooses a phase spontaneously, i.e. the ground state of a superconductor has a lower symmetry than the Hamiltonian Eq.~(\ref{eq:BCS}) which has the $U(1)$ symmetry due to phase cancellation from equal number of creation and annihilation operators.
A momentum dependent order parameter $\Delta_{\kv}^{s/t}$ then corresponds to pairing between different lattice sites. This can be seen mathematically when inserting the Fourier transformation $c_{\kv\sigma}=\frac{1}{\sqrt{N}}\sum_\mathbf {R} e^{-i\mathbf{R}\cdot\kv} c_{\mathbf{R}\sigma}$ into the mean-field Hamiltonian above. Similarly to the part of the band dispersion $\epsilon_\kv$ which originates from hopping terms, the momentum dependence of $\Delta_{\kv}^{s/t}$ will be decomposed into bond order parameters connecting neighboring lattice sites. As an example, the singlet order parameter of the form $\Delta^s_\kv=\Delta_{NN} (\cos k_x-\cos k_y)$ decomposes into NN bond order parameters with positive magnitude on horizontal bonds and negative magnitude on vertical bonds, see Fig.~\ref{fig_el_phonon_SF} (e,f).

As we have seen, the eigenstates cannot have a fixed electron number, so we introduce linear combinations of electrons and holes as new fermionic operators via a
Bogoliubov transformation
\begin{align}
\left(\begin{array}{c}
       \gamma_{\kv\up}\\
       \gamma_{-\kv\down}^\dagger
      \end{array}\right)
=
\left(\begin{array}{cc}
       u_\kv &-v_\kv\\
       v_\kv^*& u_\kv
      \end{array}\right)
\left(\begin{array}{c}
       c_{\kv\up}\\
       c^\dagger_{-\kv\down}
      \end{array}\right).\label{eq_bogo_trafo}
\end{align}
This transformation is unitary, i.e. preserves the anticommutation relations $\{\gamma_{\kv\sigma},\gamma^\dagger_{\kv'\sigma'}\}=\delta_{\kv,\kv'}\delta_{\sigma,\sigma'}$, yielding $|u_{\kv}|^2+|v_\kv|^2=1$ and the remaining freedom is used to make the Hamiltonian diagonal, i.e. the terms containing only quasiparticle annihilation operators should vanish.
This condition yields for the components of the eigenvectors
\begin{align}
  u_\kv=\sqrt{\frac 12 \left(1+\frac{\epsilon_\kv}{E_\kv}\right)},\qquad
    v_\kv=\frac{\Delta_\kv}{|\Delta_\kv|}\sqrt{\frac 12 \left(1-\frac{\epsilon_\kv}{E_\kv}\right)}
\end{align}
with the quasiparticle dispersion
\begin{equation}\label{eq_qp}
 E_\kv=\sqrt{\epsilon_\kv^2 +|\Delta_\kv^{s/t}|^2}.
\end{equation}
At this point, we note that the new eigenenergies $E_\kv$ depend only on the magnitude $|\Delta_\kv|$, while the new operators, i.e. the eigenstates also depend on the phase (or sign) of $\Delta_\kv$. Ignoring constant terms, we can rewrite the Hamiltonian as
\begin{align}
 H_{\mathrm{BCS}}=\sum_{\kv\sigma}E_\kv \gamma_{\kv\sigma}^\dagger \gamma_{\kv \sigma},
\end{align}
now expressed in the quasiparticle operators $\gamma_\kv$ describing excitations above the ground state of the superconductor. The expectation values in  Eq.~(\ref{eq_gap})  can now be calculated using the fermionic properties, Eq.~(\ref{thermal_expectation}) yielding the self-consistency condition
\begin{align}
\Delta_\kv^{s/t}=-\sum_{\kv'}V^{s/t}(\kv,\kv')\frac{\Delta_{\kv'}^{s/t}}{2E_{\kv'}}\tanh\Big(\frac{\beta E_{\kv'}}{2}\Big)\,.
\label{eq_self}
\end{align}
The dependence on the (inverse) temperature $\beta=1/(k_{\mathrm B}T)$ originates in the fermion statistics, see Sec.~\ref{thermal_exp}.

\begin{figure}[tb]
\centering
\includegraphics[width=\linewidth]{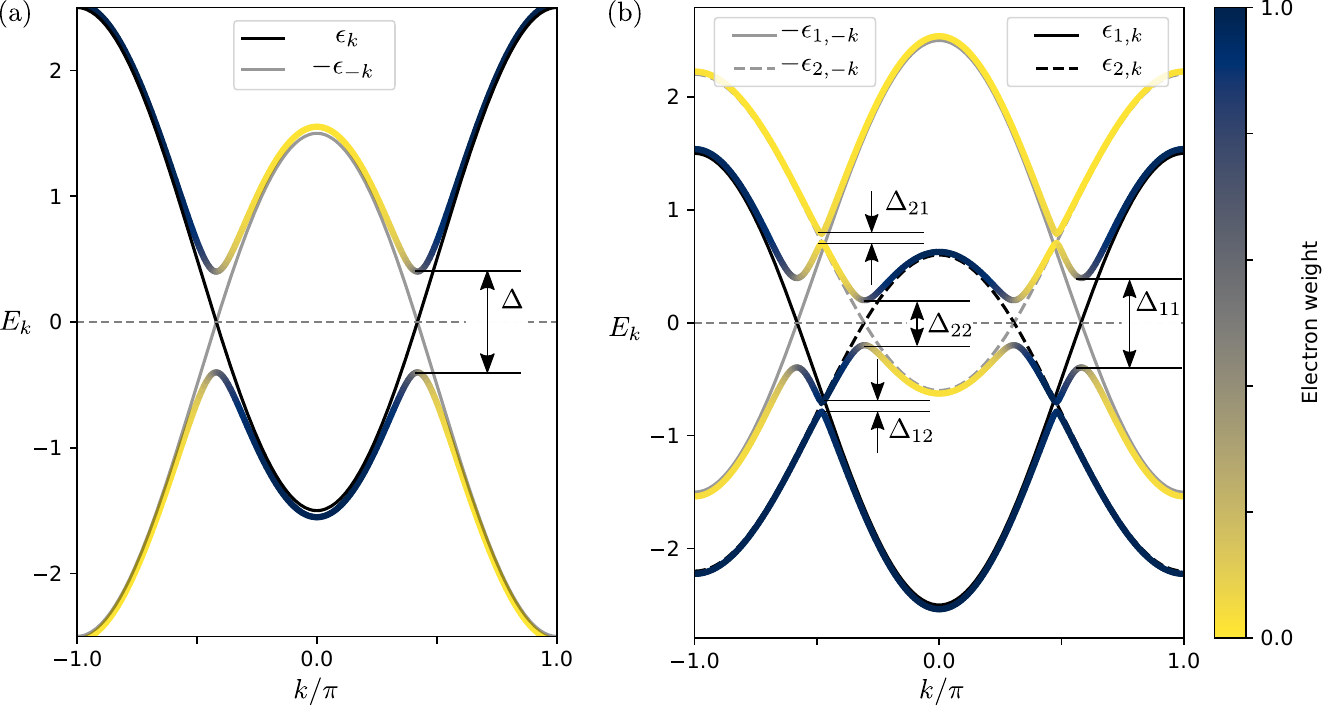}
\caption{Bogoliubov quasiparticle dispersion of (a) a single band superconductor and (b) a two band superconductor. The Bogoliubov quasiparticle dispersion $E_{\kv}$ deviates from the one in the normal state (black lines) and the dispersion of the holes (gray lines) close to the Fermi level where a gap $\Delta$ is opened. For the two-band generalization, the quasiparticle dispersions $E_{i,\kv}$ are obtained from a diagonalization of a matrix, where gap openings of different magnitudes $\Delta_{ii}$ between the same dispersions $\epsilon_{i,\kv}$ and $-\epsilon_{i,\kv}$ occur, but also at finite energy where $\epsilon_{1,\kv}$ crosses with $\epsilon_{2,\kv}$ and a gap opens from the interband pairing $\Delta_{12}$ and $\Delta_{21}=\Delta_{12}^*$. Strong interband pairing leads to a less stable superconducting ground state, can introduce Bogoliubov Fermi surfaces\cite{Setty2020} and relates to phenomena such as odd-frequency pairing\cite{Linder2019}.}
       \label{quasiparticle_dispersion}
\end{figure}

As a side remark, we note that this value of the order parameter actually minimizes the free energy $F$, i.e. the equation above can also be derived from finding a stationary point via $\frac{\delta F}{\delta \Delta_\kv}=0$~\cite{Coleman2015}. The energy is reduced compared to the metallic state once the temperature is low enough  Eq.~(\ref{eq_self}) has a nontrivial solution. In Fig.~\ref{quasiparticle_dispersion}, we visualize the quasiparticle energies $E_\kv$ where it becomes obvious that these arise from diagonalizing a $2\times 2$ matrix with the order parameter $\Delta^{s/t}_\kv$ on the off-diagonal elements. This introduces a gap opening at the Fermi level. If there are multiple bands, the gap opening occurs for each band (with different magnitude) and there are additional gap openings from \emph{interband pairing}, $\Delta_{12}$,  at higher energies, see Fig.~\ref{quasiparticle_dispersion}(b). Strong interband pairing leads to a less stable superconducting ground state, can introduce Bogoliubov Fermi surfaces\cite{Setty2020} and relates to phenomena such as odd-frequency pairing\cite{Linder2019}.

\subsection{Conventional pairing}
\label{sec_conventional}
\begin{figure}[tb]
\centering
\includegraphics[width=0.5\linewidth]{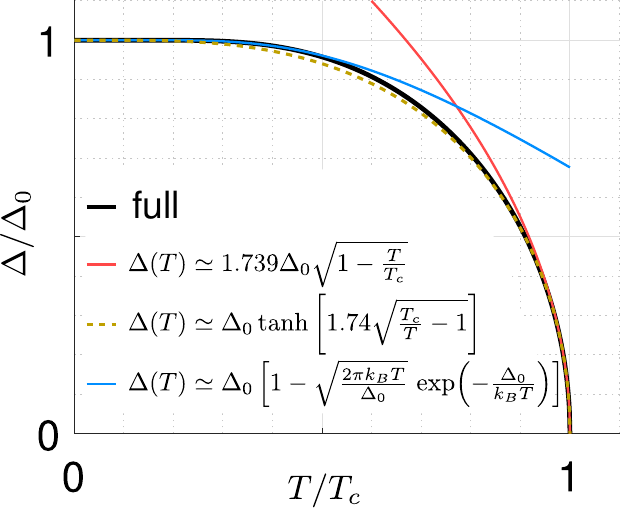}
\caption{Mean field order parameter from solution of the non-linear equation for $\Delta=\text{const.}$ in comparison to common expansions at $T=0$ and $T=T_c$ as well as used approximation over the full temperature range.}
       \label{fig_mean_field_solution_mod}
\end{figure}

Let us first look at the gap equation Eq.~(\ref{eq_self}) in the case of the conventional pairing, in order to dissect the differences towards the unconventional case. For the effective electron-phonon interaction, we assume a constant attractive interaction, i.e. $V_{\mathbf{k}\mathbf{k}'}=-V$ with $(V>0)$ within an energy window $|\epsilon_\kv|<\omega_D$ around the Fermi energy \cite{Coleman2015}. Inserting this into our equation, it simplifies to
\begin{align}
\Delta_\kv=V\sum_{\kv'; |\epsilon_{\kv'}|<\omega_D}\frac{\Delta_{\kv'}}{2E_{\kv'}}\tanh\Big(\frac{\beta E_{\kv'}}{2}\Big)\,,
\label{eq_self1}
\end{align}
where we have dropped the singlet superscript $s$.
The r.h.s. does not depend on $\kv$, so the order parameter is just a constant, $\Delta_{\mathbf{k}}=\Delta$.
Assuming a nonzero $\Delta$, we can divide by it and obtain
\begin{equation}1=V\sum_{\kv'; |\epsilon_{\kv'}|<\omega_D}\frac{1}{2E_{\kv'}}\tanh\Big(\frac{\beta E_{\kv'}}{2}\Big)\,.
\end{equation}
Looking at the terms under the momentum sum, one sees that the integrand goes to zero when $E_\kv$ is large. In detail, the integrand grows for lower temperatures with a maximum at the Fermi surface where $E_\kv$ is smallest. The temperature dependence of the constant $s$-wave solution is shown in Fig.~\ref{fig_mean_field_solution_mod} together with some approximate analytical formulas used in the literature.

\subsection{Momentum-dependent interaction and unconventional pairing}

When the interaction is considered to be momentum-dependent, it cannot be factored out of the sum any more. As argued above, the superconducting pairing has the largest contribution from states close to the Fermi level, so we restrict our discussion to the order parameter at the Fermi level for now.

Considering an electronic structure with parabolic bands, which is also a good approximation in the case of small (or large electron density) of the mentioned tight-binding model, Eq.~(\ref{eq_disp_single}), we use $\epsilon_\kv=\frac{\kv^2}{2m_\mathrm{eff}}$ with some effective mass $m_\mathrm{eff}$. In this case, the Fermi surface is just a sphere with radius $k_F=\sqrt{2m_\mathrm{eff} \mu}$. We note at this point that natural units are used throughout this work, i.e. the Planck constant is set to unity, $\hbar=1$ such that the mass  $m_\mathrm{eff}$, the momentum $\mathbf k$ have the dimension of energy. It is also convenient to set $k_\mathrm{B}=1$ so that the temperature also has the dimension of energy. With the picture in mind that superconductivity is governed by low energy properties at the Fermi surface, we define the superconducting order parameter $\Delta_{\kv_F}$ on a sphere of radius $\kv_F$. Recalling that spherically symmetric problems, for example boundary value problems in electrodynamics or the Schrödinger equation for an electron in the hydrogen atom, can be solved with a separation ansatz where the angular part of the solution function can be expanded in the (real-valued) spherical harmonics $Y_l^m(\theta,\phi)$, we do the same with the order parameter,
\begin{equation}
\Delta_{\kv_F}^{s/t} =\sum_{m,l} \Delta_{lm} Y_l^m(\theta,\phi).
\end{equation}
Here, $\Delta_{lm}$ are expansion coefficients. The interaction as a function of two momenta $\kv$ and $\kv'$ (on the Fermi surface) exhibits a similar expansion with products of spherical harmonics of the two pairs of angles $(\theta,\phi)$ and $(\theta',\phi')$,
\begin{equation}
V^{s/t}(\kv,\kv')=\sum_{m,l} V_l  Y_l^m(\theta,\phi) Y_l^m(\theta',\phi').\label{expansion_Vkk}
\end{equation}
Orthogonality of the basis functions decouples the gap equation Eq.~(\ref{eq_self}) and it can be solved for the critical temperature $T_c=1.13\omega_l \exp[-\frac{1}{V_l N(0)}]$ for each $l$ similar to the original BCS result\cite{BCS1957}.
The order parameter in angular channel $l$ with the largest $T_c$ is then realized. The singlet states correspond to even $l$ and the triplet states to odd $l$. In analogy to the nomenclature of the electronic states of the hydrogen atom, one assigns the notation $s$-wave pairing for $l=0$, $p$-wave pairing to $l=1$, $d$-wave pairing to $l=2$ etc. In this spherically symmetric case, the $l=0$ solution transforms trivially under all rotations, $l=1$ as the $p_x$ or $p_y$ atomic orbital, $l=2$ as the $d_{xy}, d_{x^2-y^2}, d_{xz}, d_{yz}$ or $d_{z^2}$ orbital.

\subsubsection{Order parameter classifications in crystalline systems}
In crystalline systems, the full rotational symmetry is broken down to discrete rotation or mirror symmetries, so the classification or labeling of order parameters (and other quantities such as electronic bands, phonon modes) is done using these symmetry operations. Order parameters transform in a certain way under the symmetry operations of the system as can be deduced from group theory \cite{Ramires2022}. At this point, we do not want to give an introduction to group theory, but instead present the structure of the point group $D_4$ as an example. This describes the square lattice as shown in Fig.~\ref{fig_real_reciprocal}(a) which is a relevant example for a number of quasi-two dimensional unconventional superconducting materials with tetragonal symmetry such as heavy fermion materials\cite{Steglich1979}, cuprates\cite{Scalapino2012}, iron-based superconductors\cite{Stewart2011}, nickelates\cite{Norman2020} and Sr$_2$RuO$_4$\cite{Maeno2024}.

The square lattice has the following 8 symmetry elements: 4 rotations: $E$, $C_4$, $C_4^2$, $C_4^3$ describing the identity, and rotations by 90$^\circ$ and its multiples. There are two pairs of mirror operations $\sigma_x$ and $\sigma_y$, the mirrors at the two coordinate axes, and  $\sigma_d$ and $\sigma_{d'}$, the mirror operations along the diagonals.

Using the multiplication relations of the group elements, one can find the number of irreducible representations (irreps), in our case 5 \cite{Ramires2022}. Each irrep is then associated with a character describing how the irrep transforms under the operation. The one dimensional irreps (with one element) are either invariant, character $1$, or pick up a minus sign ($-1$) under the symmetry operation. The elements of the two-dimensional irrep $E$ can be invariant (+2), each picks up a minus sign ($-2$) or transform into each other (0). Order parameters of superconducting states that do not break crystalline symmetries correspond to one of the irreps, thus the corresponding value of the character tells under which transformations a sign change is forced which fixes the position where the order parameter has to be zero by symmetry (nodes).

\begin{table}[tb]
\centering
\begin{tabular}{|c|c|c|c|c|c|c|}
\hline
irrep & $E$ & $2C_4$ & $C_2$ & $2\sigma_v$ & $2\sigma_d$& name \\
\hline
$A_1$ & $1$ & 1 & 1 & 1 & 1&$s$\\
\hline
$A_2$ & 1 & 1 & 1 &$ -1$ & $-1$&$g$ \\
\hline
$B_1$ & 1 & $-1$ & 1 & 1 & $-1$&$d_{x^2-y^2}$ \\
\hline
$B_2$ & 1 & $-1$ &$ 1$ & $-1$ & 1&$d_{xy}$ \\
\hline
$E$ & 2 & 0 & $-2$ & 0 & 0&$(p_x,p_y)$ \\
\hline
\end{tabular}
\caption{Character table for the $C_{4v}$ point group with groups of symmetry elements and the often associated name of the irrep.}
\label{tab:c4v_character_table}
\end{table}

When describing the superconducting order parameter $\Delta_\kv$ on the Fermi surface (or in the full Brillouin zone), it is clear that a complete set of functions (in two or three dimensions) is needed and therefore each irreducible representation contains infinitely many basis functions that all transform in the same way.
For example, the trivial irrep A$_1$ which is invariant under all symmetry operators can be expanded as
\begin{equation}
 \Delta_\kv=\Delta_0 + \Delta_{NN} (\cos k_x + \cos k_y)+ \Delta_{3N} (\cos 2k_x + \cos 2k_y)+\cdots.
\end{equation}

{\bf Accidental nodes} For the trivial irrep, there are two possibilities: If the first coefficient $\Delta_0$ is large and dominant such that the order parameter is finite everywhere in the Brillouin zone and the system remains fully gapped. If the other coefficients are large, there is the possibility that the order parameter exhibits nodes, $\Delta_\kv=0$, which are accidental and yield a $E_\kv=0$ if the nodes intersect the Fermi surface.

{\bf Symmetry enforced nodes}
For the nontrivial irreps, it is clear that the minus sign anywhere in the character table forces a sign change of the order parameter, which necessarily introduces a nodal line (or nodal plane in three spatial dimensions) at the momenta which are invariant under such an operation. For example, the $B_1$ irrep (alias $d_{x^2-y^2}$) has character $-1$ under the two mirror operations along the diagonal $\sigma_d$, see Table~\ref{tab:c4v_character_table}. The line $k_x=k_y$ is unchanged under such a mirror, the order parameter has to change sign, so only the value $0$ is allowed, and the order parameter has a node. Still, there is the possibility that the nodes do not intersect with the Fermi surface and no $\kv$ exists where  $E_\kv=0$, i.e. also nontrivial pairing can be fully gapped. This would occur for the  $B_1$ irrep on the Fermi surface shown in panel (b) of Fig.~\ref{fig_lifshitz} where the lines $k_x=\pm k_y$ do not intersect with the Fermi surface.

\begin{figure}[tb]
\centering
 \includegraphics[width=0.5\linewidth]{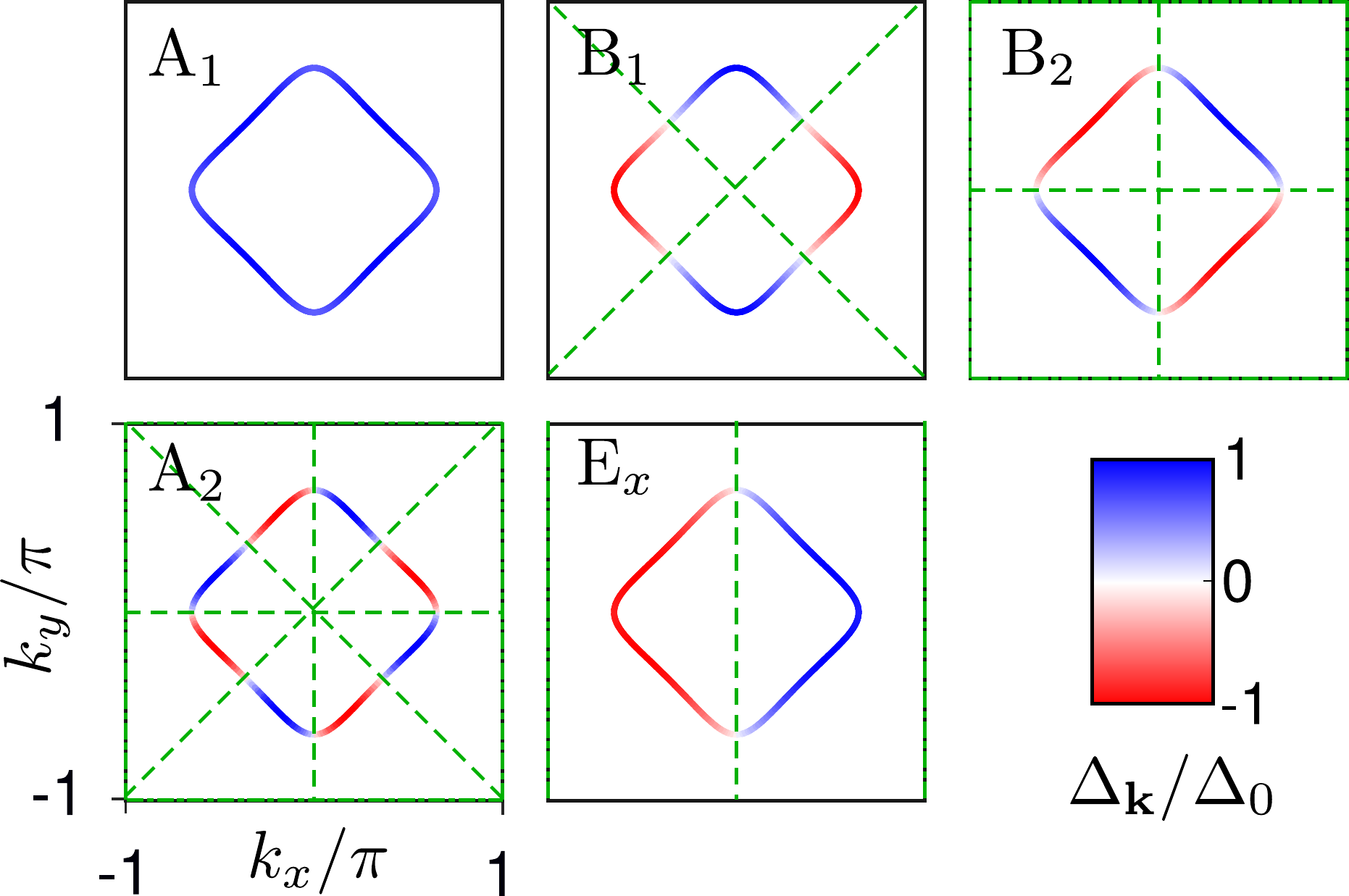}
\caption{Lowest order harmonics of the 5 different irreps on the square lattice, see table \ref{tab:c4v_character_table} plotted on a Fermi surface with only one $\Gamma$ centered pocket, but lattice effects, see Fig.~\ref{fig_lifshitz}. The order parameters exhibit sign changes and symmetry enforced nodal lines (green dashed) as well as additional nodal lines (green, dash-dotted) due to the periodicity of the Brillouin zone which is indicated by a black frame.}
\label{fig_trivial_irrep}
\end{figure}

Relevant order parameters for the square lattice are presented in Fig.~\ref{fig_trivial_irrep} with a Fermi surface centered at the $\Gamma$ point that intersects with nodal lines of all nontrivial order parameters to illustrate the symmetry-enforced nodal lines that can be understood from the entries in Table~\ref{tab:c4v_character_table}.

{\bf Chiral order parameters}
For two dimensional irreps (or in the case of accidental degeneracy where the critical temperature for two components is identical), there is the possibility to form a complex linear combination, a so-called chiral order parameter\cite{Ramires2022}. The overall magnitude $|\Delta_\kv|$ never goes to zero, but the order parameter still acquires a minus sign under some transformations. An example would be the complex order parameter $\Delta_\kv=\Delta_0 (\sin k_x + i\sin k_y)$ with  $|\Delta_\kv|\neq 0$ everywhere but still acquires a sign change under the 180$^\circ$ rotation ($C_2$).

\subsubsection{Example: Pairing from $t-J$ model}
\begin{figure}[tb]
\centering
 \includegraphics[width=\linewidth]{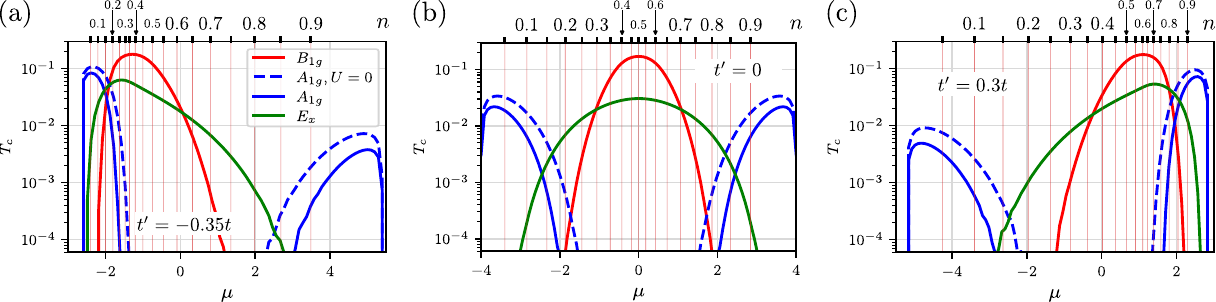}
\caption{Critical temperature $T_c$ as calculated from the linearized gap equation for a model of electrons on a square lattice subject to the interaction in Eq.~(\ref{eq_tJ}) of the $t-J$ model with $U=0.5t$, $J=1.5t$. We show $T_c$ on a logarithmic scale as function of chemical potential $\mu$ and filling per spin $n$ (upper horizontal axis). The phase diagram with only nearest-neighbor hopping $t$ is symmetric around $\mu=0$ (b) while the introduction of the symmetry-allowed finite next-nearest-neighbor hopping $t'$ breaks this symmetry for $t'=-0.35t$ (a) and $t'=0.3t$ (c). Only the $A_{1g}$ instability is suppressed from repulsive onsite interaction $U$ as evidenced by the calculation without it.}
\label{fig_LGE}
\end{figure}

To illustrate the discussion about the irreducible representations, basis functions and multiple critical temperatures, let us take the pairing interaction from the $t-J$ model,
\begin{align}
 V(\kv,\kv')=U-\frac{3J}2 \left[\cos(k_x-k_x')+\cos(k_y-k_y')\right]\label{eq_tJ}
\end{align}
which is obtained from augmenting the Hubbard model with nearest neighbor Heisenberg Hamiltonian $H_\text{int}=J\sum_{\langle i,j\rangle}\vec S_i\cdot \vec S_j$ and rewriting the relevant terms
in the form of Eq.~(\ref{eq:BCS})\cite{Coleman2015}. Let us ask the question which superconducting order parameter is realized in this model. For this, we solve the gap equation Eq.~(\ref{eq_self})
and calculate the $T_c$ for each channel. Some technical details are given in Sec.~\ref{sec_lge_details} which illustrate the method and produce the results in Fig.~\ref{fig_LGE} as answer to the question about the superconducting order parameter. Depending on the choice of the chemical potential $\mu$ or the number of electrons (upper axis), different states have the largest $T_c$. The reason lies in the gap equation that has dominant contribution from the states close to the Fermi level. In Fig.~\ref{fig_lifshitz}, we have seen that our model can have Fermi surfaces at different places in the Brillouin zone where the preferred order parameter can better gap out the electronic structure. The overall temperature scale follows trends from the density of states as can be seen comparing to Fig.~\ref{single_band_lifshitz}(b). The presence of the repulsive onsite interaction $U$ suppresses the trivial irrep A$_1$ ($s$-wave), but leaves the compensated nontrivial order parameters unchanged.

\subsubsection{Solutions without projecting into irreps}
The simple interaction in Eq.~(\ref{eq_tJ}) contains only onsite $U$ and NN terms $J$. More generally, the interaction is a function of $\kv$ and $\kv'$ which has all Fourier components, i.e. bond interactions. In this case, the leading superconducting instability at $T_c$ can still be classified by the irreducible representation, but it does exhibit higher harmonics and display a more complex momentum dependence. A strategy to solve the linearized gap equation can be to expand the interaction in bonds or lattice harmonics with a truncation and solve the corresponding eigenvalue equations, or perform a momentum space discretization and solve the eigenvalue problem in the Brillouin zone
\begin{equation}
 \Delta_\kv^{s/t} =\sum_{\kv'}  M^{s/t}(\kv,\kv') \Delta_{\kv'}^{s/t}
\end{equation}
i.e. calculate eigenvectors and eigenvalues of the matrix
\begin{equation}
 M^{s/t}(\kv,\kv')=-\frac{V^{s/t}(\kv,\kv')} {2\epsilon_{\kv'}}\tanh\Bigl(\frac{\beta \epsilon_{\kv'}}{2}\Bigr),\label{eq_def_M}
\end{equation}
and find the temperature at which the largest eigenvalue reaches unity.
We note that this matrix is not symmetric since on the r.h.s. the weight $[2\epsilon_{\kv'}]^{-1}\tanh(\frac{\beta \epsilon_{\kv'}}{2})$ appears only in $\kv'$ and not in $\kv$. So, one might wonder whether the matrix has real eigenvalues (and real-valued eigenvectors) in general.
This is indeed the case, as one can see from the following: Let us define a positive diagonal matrix $\Lambda^2_{\kv'}=[2\epsilon_{\kv'}]^{-1}\tanh(\frac{\beta \epsilon_{\kv'}}{2})$. We can now write Eq.~(\ref{eq_def_M}) as matrix multiplication, $M_{\kv,\kv'}=-V_{\kv,\kv'}\Lambda^2_{\kv'}$. Next, we multiply this equation with the diagonal matrices $\Lambda_{\kv}$ and ${\Lambda_{\kv'}}^{-1}$ from left and right to obtain ${\Lambda_\kv} M_{\kv,\kv'}{\Lambda_{\kv'}}^{-1}=- {\Lambda_{\kv}}V_{\kv,\kv'} {\Lambda_{\kv'}}\equiv \tilde M$.
This is now a symmetric (real-valued) matrix that has real eigenvalues $\lambda_i$ and eigenvectors $\tilde g_i$ obeying $\tilde M\tilde g_i=\lambda_i\tilde g_i$. Substituting the definition of $\tilde M$ and multiplication with ${\Lambda_{\kv}}^{-1}$ yields
$M_{\kv,\kv'} {\Lambda_{\kv'}}^{-1}\tilde g_i=\lambda_i {\Lambda_{\kv}}^{-1}\tilde g_i$,
in other words $g_i\equiv {\Lambda_{\kv}}^{-1}\tilde g_i$
is eigenvector to the original matrix in Eq.~(\ref{eq_def_M}) with eigenvalue $\lambda_i$.

\subsubsection{Weak-coupling approximation}
Similarly to the discussion on momentum independent interactions, we can constrain the interactions to be only on an energy shell of $\omega_D$ around the Fermi level, split up the summation along and perpendicular to the Fermi surface, and solve the integral perpendicular to the Fermi surface. Then, the mathematical problem is cast to solving the equation \cite{HirschfeldCRAS}
\begin{align}
   -\frac{1}{V_G}\int_{FS} dS' V^{s/t}(\kv,\kv')\frac{g_i(\kv')}{|v_F(\kv')|}
  =\lambda_i g_i(\kv)\,.\label{eq_lin_gap}
\end{align}
for the largest eigenvalues $\lambda_i$.
Here $V_G$ is the volume of the Brillouin zone, the integral $\int_{FS}$ is over the Fermi surface evaluated at the points $\kv'$, $g_i(\kv)$ is the gap symmetry function which contains the momentum dependence of the superconducting instability with eigenvalue $\lambda_i$ where the instability with the largest eigenvalue is realized at $T_c$. Formally, the critical temperature can be calculated from the eigenvalue by $T_c=1.13 \omega_D e^{-\frac{1}{\lambda_i}}$. In many cases, the energy scale $\omega_D$ from the effective pairing interaction is difficult to estimate, so critical temperatures cannot be calculated in a controlled way. Again, a discretization of the Fermi surface into a set of Fermi points $\kv'$ with associated area $l_{\kv'}$ casts the problem of finding the eigenvalues and eigenvectors of the matrix
\begin{eqnarray}
  M^{s/t}_{\kv,\kv'}=-\frac{1}{V_G}\frac{l_{\kv'}}{|v_F(\kv')|} V^{s/t}(\kv,\kv'),
  \label{eq:LGEmatrix}
\end{eqnarray}
which has by the same arguments from above only real eigenvalues.

To summarize some findings, we can state that repulsive Coulomb interactions tend to generate pairing states that are compensated, eventually leading to a sign change. 
Even when the interactions are repulsive for all combinations of momenta $\kv$ and $\kv'$, there is a solution to the gap equation, Eq.~(\ref{eq_self}) if there is a sign change of the order parameter. In this case, the order parameter is large and has the opposite sign for combinations $\kv$ and $\kv'$ where the interaction is most repulsive. This leads to a guiding principle: Nodes appear on parts of the Fermi surface where interactions are weak, gap maxima with opposite sign appear on parts of the Fermi surface where the interactions are strong and repulsive. A trivial irrep $A_1$ can be a solution in this case if the Fermi surface sustains it, but these states have nodal lines (that may or may not cross the Fermi level). For the triplet channel, the antisymmetrization as given in Eq.~(\ref{eq_pairing_symm}) results in a sign change of the pairing interaction, thus it is formally attractive $ V^{t}(\kv,\kv')<0$, but the order parameter is odd parity $\Delta_\kv^t=-\Delta_{-\kv}^t$, such that it has to exhibit nodes somewhere in the Brillouin zone (at the $\Gamma$ point as inversion invariant point) and often at the Brillouin zone boundary because of the periodicity, see Fig.~\ref{fig_trivial_irrep}. Finally, optimization of the momentum dependence to lower the total energy can lead to accidental nodes (in addition to the symmetry enforced ones) even for nontrivial irreps\cite{Romer2015}.

\subsection{Behavior of the quasiparticle excitations}

Thermodynamic properties and spectroscopic probes in superconductors are governed by the quasiparticle excitations; their energies have been calculated in Eq.~(\ref{eq_qp}) and only depend on the magnitude of the superconducting order parameter. However, the quasiparticle states do depend on the sign (or complex phase) of the order parameter as seen from Eq.~(\ref{eq_bogo_trafo}) leading to $\gamma_{\kv\up}=u_\kv c_{\kv\up}-v_\kv c^\dagger_{-\kv\down}$ where the function $v_\kv$ depends on the phase.

\begin{figure}[tb]
\centering
 \includegraphics[width=\linewidth]{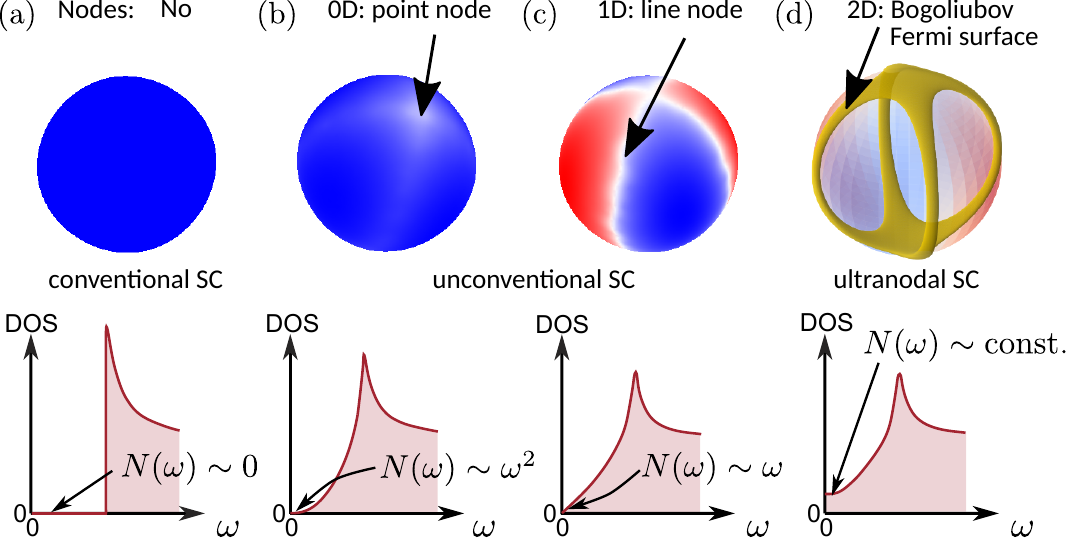}
\caption{Density of states of 3D superconductors: (a) The conventional superconductor is fully gapped on the Fermi surface (blue sphere) and exhibits no states at low energy as seen in the density of states (DOS) below. (b,c) Unconventional superconductors can exhibit point nodes (zero dimensional, arrow) or line nodes (one dimensional, white line), where symmetry imposes the sign change of the order parameter (red/blue). The density of states at low energies exhibits characteristic power laws. (d) Example of an ultranodal superconductor (here a $d$-wave superconductor in a magnetic field\cite{Setty2020}). There are contours of zero energy that form two dimensional Bogoliubov Fermi surfaces (yellow), such that the density of states at zero energy remains constant\cite{Setty2020,Agterberg2017}.}
\label{fig_dos_SC}
\end{figure}

First, we discuss the properties of the excitation gap, i.e. $E_\kv$ and contrast its behavior between trivial and nontrivial pairing states. Representative cases for gapping in three spatial dimensions are shown schematically in Fig.~\ref{fig_dos_SC}.

 {\bf Full gap:} If the order parameter is a constant $\Delta_\kv=\Delta$, we observe that $E_\kv=\sqrt{\epsilon_\kv^2 +|\Delta|^2}$ has its minimum at $|\Delta|$, i.e. there are no quasiparticle excitations below that scale, the superconductor is fully gapped. A variant of this case occurs if the order parameter is not constant, but remains finite at all $\kv_F$ where $\epsilon_\kv=0$, also in this case there are no states at energies lower than $\Delta_{\text{min}}$.
 
{\bf   Nodal gap:} A nontrivial superconducting order parameter typically exhibits high symmetry lines (or planes in three dimensional systems) where the order parameter vanishes, $\Delta_{\kv_{\text{node}}}=0$. For accidental nodes (in the trivial or nontrivial irreps), this might happen at curved lines (or surfaces). If these lines (or planes) intersect the Fermi surface, we have k-points where $\epsilon_\kv=0$ and $\Delta_{\kv}=0$, so $E_\kv=0$. This happens on a manifold that has a lower dimension than the Fermi surface, and the system is classified as a nodal superconductor. For multiband systems, the quasiparticle energies and eigenstates are obtained from a matrix diagonalization such that there is not a closed form expression as Eq.~(\ref{eq_qp}) and the nodes might appear away from the Fermi surface\cite{Chubukov2016}.

{\bf   Inflated gap:} In systems that break time-reversal symmetry, there is a third possibility where $E_\kv=0$ on a manifold that has the same dimension as the Fermi surface. This ultranodal state exhibits properties of a superconductor and a normal metal. It has finite order parameter and pair correlation as given in Eq.~(\ref{eq_pair_corr}), it is however not gapped in the same sense, and has remaining quasiparticle excitations at zero energy comparable to the normal state. It can occur in systems where electronic states of higher angular momentum pair\cite{Agterberg2017} or from the interplay of multiple order parameters in multiband systems \cite{Setty2020}.

\subsection{Density of states of a superconductor}

The gapping out of states at low energies is one hallmark of superconductivity and is commonly probed experimentally to find evidence for superconductivity in a material and constrain the superconducting order parameter by measuring details of the density of states. This can be done, for example, by scanning tunneling experiments as a more direct probe of the density of states\cite{Hoffman2011}, or by probing quasiparticle excitations in thermodynamic quantities. Understanding the connection between the density of states and the superconducting order parameter is crucial to correctly interpret experimental data.

\paragraph*{Conventional superconductor:} 
Let us first recall the properties of a conventional superconductor with a constant order parameter. The excitation energy is given by Eq.~(\ref{eq_qp}), i.e. $E_\kv=\sqrt{\epsilon_\kv^2+|\Delta|^2}$, i.e. there are no states between $-|\Delta|<\omega<|\Delta|$. An evaluation of Eq.~(\ref{eq_dos}) with this quasiparticle dispersion yields the result
\begin{equation}
 N(\omega)=N_0\frac{|\omega|}{\sqrt{\omega^2-|\Delta|^2}}
\end{equation}
for energies $|\omega|>|\Delta|$, i.e. it exhibits a divergence right at the energy scale $\Delta$, usually referred to as coherence peak.

\paragraph*{Unconventional superconductor:}
The key difference of unconventional superconductivity is that the order parameter $\Delta_\kv$ is momentum dependent and typically has (1) minima (2) maxima and often (3) nodes. To discuss the density of states in these three cases, we expand Eq.~(\ref{eq_qp}) in momentum space at selected points and to calculate the behavior of the density of states.

\begin{figure}[tb]
\centering
 \includegraphics[width=\linewidth]{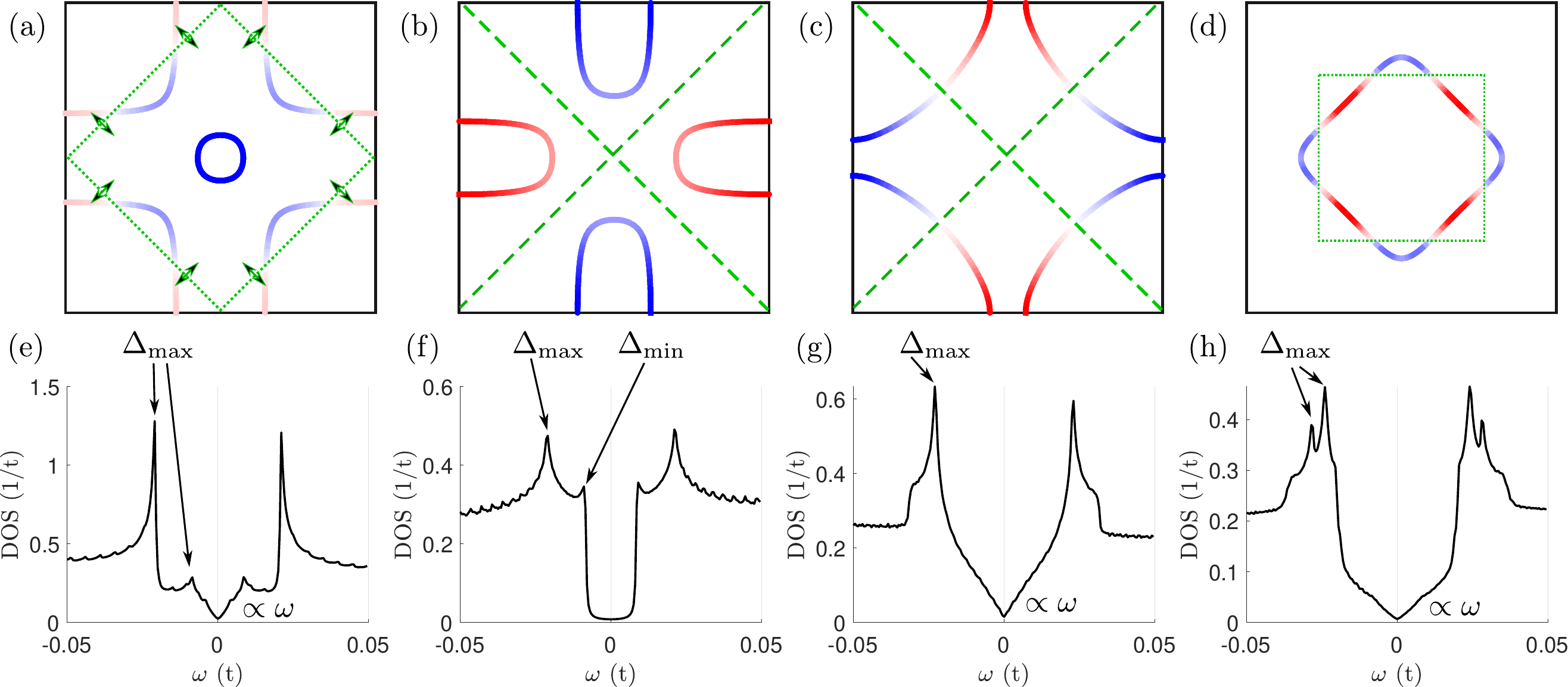}
\caption{Order parameters on Fermi surfaces of the square lattice model of Eq.~(\ref{eq_disp_single}). (a) Lowest order A$_1$ state with accidental nodes crossing the Fermi surface (dotted lines) that can move freely (arrows). (b) Fully gapped B$_1$ state where the gap nodes do not intersect with the Fermi surface. (c) Order parameter as calculated from Eq.~(\ref{eq:LGEmatrix}) together with Eq.~(\ref{eq_sf_pairing_vertex}) within the spin-fluctuation approach. (d) Leading A$_1$ instability from the same approach which exhibits higher order harmonics such that accidental nodes are on the Fermi surface. (e-h) show the relevant electronic density of states, Eq.~(\ref{eq_dos_SC}).}
\label{nontrivial_gaps_single_band}
\end{figure}

\paragraph*{Minimum or maximum of $E_\kv$}
Let us start with the extrema of the order parameter where the quasiparticle dispersion increases quadratically in the direction(s) parallel $\qv_{\parallel}$ and in the direction perpendicular $q_\perp$ to the Fermi surface, thus the expansion around $\kv_{\text{ext}}$ becomes
\begin{equation}
 E_\kv\approx \Delta_{\text{ext}}+\frac{v_F^2 q_{\perp}^2}{2 \Delta_{\text{ext}}}+\sum_i \frac{\kappa_i}{2}q^2_{i,\parallel}\label{eq_expansion}
\end{equation}
where $v_F$ is the Fermi velocity at the expansion point and $\kappa_i=\partial_{q_{i,\parallel}}^2 \Delta_{\kv_{\text{ext}}}$ are the order parameter masses, i.e. the eigenvalues of the Hessian matrix. Next, we analyze the corresponding contributions to the density of states from expansions of the dispersion around singular points \cite{Hove1953,  Classen2020}.
For a gap minimum, we have a quadratic quasiparticle dispersion with anisotropic masses; rescaling the momentum variables, this casts back to the calculation of the density of states in a parabolic band with the result
\begin{equation}
 N(\omega)= \Theta(\omega-\Delta_{\text{min}})
 \left\{\begin{array}{ll}
                  \frac{\sqrt{\Delta_{\text{min}}}}{2\pi v_F\sqrt{\kappa}}& D=2\\
                   \frac{\sqrt{2\Delta_{\text{min}}}}{\pi^2 v_F\sqrt{\kappa_1 \kappa_2}}\sqrt{\omega -\Delta_{\text{min}}}& D=3
                  \end{array}
\right.
\end{equation}
for each of the minima, i.e. there is a jump or onset of square root behavior. These features can be seen in the calculated density of states in Fig.~\ref{nontrivial_gaps_single_band} (f).

\paragraph*{Saddle point of $E_\kv$}
For a maximum of the order parameter on the Fermi surface, $\kappa_i<0$ and we obtain an expansion of the quasiparticle energy that exhibits a saddle point, i.e. $E_\kv$ goes down in one direction and increases in others. In this case we obtain a van Hove singularity with the additional singular contribution
\begin{equation}
  N(\omega)=\Theta(\omega-\Delta_{\text{max}}) \frac{\sqrt{2\Delta_{\text{max}}}}{2\pi^2v_F\sqrt{|\kappa_1||\kappa_2|}}\sqrt{\omega-\Delta_{\text{max}}}\quad D=3
\end{equation}
in three dimensions. The fine-tuned case of a saddle point of the order parameter will give a negative contribution of the same magnitude below the saddle point value $\Delta_s$, $N(\omega)\sim -\Theta(\Delta_s-\omega)\sqrt{\Delta_s-\omega}$. In two dimensions, the order parameter maximum produces a logarithmic divergence of the density of states of the form
\begin{equation}
  N(\omega)=\frac{ \sqrt{\Delta_{\text{max}}}}{2\pi^2v_F \sqrt{\kappa}}\ln \left(\frac{\Lambda}{|\omega-\Delta_{\text{max}}|}\right)\quad D=2
\end{equation}
where $\Lambda$ is the energy scale where the quadratic expansion, Eq.~(\ref{eq_expansion}) breaks down. These peaks are clearly seen at the energy of maxima in $\Delta_\kv$ in all calculated density of states in Fig.~\ref{nontrivial_gaps_single_band}.

\paragraph*{Nodes of $E_\kv$}
Finally, let us discuss the important part for the low energy density of states that is relevant for the low temperature behavior of experimentally accessible quantities such as specific heat, thermal conductivity and penetration depth of the magnetic field $\lambda$.

For the nodal points, we can use the fact that the quasiparticle dispersion is an anisotropic Dirac cone
$ E_\kv=\sqrt{v_F^2 q_\perp^2+v_\Delta^2 q_\parallel^2}$
with $v_F$ the Fermi velocity at the nodal point and $v_\Delta=\partial_{q_\parallel} \Delta_\kv$ the gap velocity. The density of states can be calculated analytically by rescaling of the momenta and evaluation of the $\delta$ function in Eq.~(\ref{eq_dos}) in polar coordinates with the result
\begin{equation}
 N(\omega)=\frac{\omega}{2\pi v_F v_\Delta}\qquad D=2,3.
\end{equation}
The linear behavior at low energies is universal, its slope depends on details as also visible in Fig.~\ref{nontrivial_gaps_single_band}(e,g,h).
The same result is obtained for line nodes in three dimensions where $v_F$ and $v_\Delta$ need to be averaged along the line node.

A vanishing order parameter on single points on a three dimensional Fermi surface yields an anisotropic Dirac cone in three dimensions when expanding around these points,
 $E_\kv=\sqrt{v_F^2 q_\perp^2+v_{1,\Delta}^2 q_{1,\parallel}^2+v_{2,\Delta}^2 q_{2,\parallel}^2}$
with the result for the density of states of
\begin{equation}
 N(\omega)=\frac{\omega^2}{2\pi^2 v_Fv_{1,\Delta}v_{2,\Delta}}\qquad D=3.
\end{equation}
Again, the power law is universal, the prefactor not.

\paragraph*{Electron density of states}
Considering the Bogoliubov transformation again, Eq.~(\ref{eq_bogo_trafo}), we see that electron tunneling as measured in  (scanning) tunneling experiments actually couples to the electron part of the Cooper wavefunction only.
Differential conductance of electron tunneling\cite{Hoffman2011} is therefore proportional to the electron density of states
\begin{equation}
 N_{\text{el}}(\omega)=\sum_\kv\left( |u_\kv|^2 \delta(\omega-E_\kv)+|v_\kv|^2 \delta(\omega+E_\kv)\right)\,. \label{eq_dos_SC}
\end{equation}
It is not symmetric with respect to $\omega \rightarrow -\omega$ and reduces to the normal state density of states when $\Delta_\kv=0$. Still, it only depends on the magnitude of $|\Delta_\kv|$ not on its phase, so different symmetries of the order parameter can give rise to the same electronic density of states as well.
In Fig.~\ref{nontrivial_gaps_single_band} some examples of order parameters on the square lattice tight binding model are shown together with calculated density of states for these nontrivial states. These reveal the features coming from gap maxima, gap minima and nodes.

We have considered here the clean limit of the density of states which might be a good approximation for a number of elementary superconductors such as Hg or $\beta$-Sn. Clean systems that show very small intrinsic scattering rates from normal state resistivity measurements can also be reached for some unconventional systems, for example Sr$_2$RuO$_4$ or UTe$_2$ where the residual resistivity ratio (RRR) can reach values of 1000. Disorder in many unconventional systems leads to suppression of critical temperature, modifications of the superconducting order parameter, introduction of bound states in the superconducting state, and changes of the density of states at low energies. In some cases these modifications inhibit a clear identification of the nature of the nodes in the order parameter, in other cases a systematic study of evolution of the density of states with disorder can be used to deduce the order parameter \cite{HirschfeldCRAS}. 

\section{Microscopic pairing interactions}

In this section, we want to summarize some concepts of microscopic pairing mechanisms and describe guiding principles to understand which symmetries of the superconducting order parameter are expected. Indeed, this topic is current research with many facets and theoretical frameworks, so we want to restrict here to the weak-coupling perspective where the low-energy electronic states, i.e. the Fermi surface, play a key role. The interested reader is referred to the reviews in Refs.~\cite{Varma_2012,Marsiglio2020} for considerations outside this paradigm.

The key idea for spin-fluctuation driven pairing interaction is already sketched in Fig.~\ref{fig_el_phonon_SF}: Due to strong interactions between electrons in materials together with the Pauli exclusion principle, the spins of the electrons do not behave as independent degrees of freedom. The spins fluctuate collectively such that the sea of neighboring spins anti-align to screen the fluctuating spin of the first electron. This effect is usually strong if there is a magnetic ordering (in many materials antiferromagnetic ordering) close by. The key properties of this interaction are that it is (1) momentum-dependent in a specific form related to the magnetic ordering and (2) it is generically repulsive $V(\kv,\kv')>0$. The latter means that we can only achieve a solution to the gap equation, Eq.~(\ref{eq_self}), in the weak-coupling approximation by exhibiting a sign change of the order parameter.

Still, the symmetry of the order parameter is not determined by the type of microscopic pairing mechanism: It is still possible to achieve trivial ($A_{1g}$) pairing states as discussed in iron-based superconductors \cite{Mazin2009,Chubukov2012} or nontrivial $B_{1g}$ (alias $d$-wave) pairing states that are relevant for cuprates and some heavy fermion systems \cite{Scalapino2012}. As a basic modeling, one can start from a local moment picture by writing down a Heisenberg model $H_\text{int}=J\sum_{ i,j}\vec S_i\cdot \vec S_j$ as discussed for example in Ref.~\cite{davis13}, or start from an itinerant model where free electrons with dispersion $\epsilon_\kv$ are additionally subject to an onsite Hubbard repulsion of strength $U$,
\begin{equation}
 H=\sum_{\kv,\sigma}\epsilon_{\kv} c_{\kv\sigma}^\dagger c_{\kv\sigma}+U\sum_{\mathbf r} c_{\mathbf r\uparrow}^\dagger c_{\mathbf r\uparrow} c_{\mathbf r\downarrow}^\dagger c_{\mathbf r\downarrow}\,.\label{eq_Hubbard}
\end{equation}

\subsection{Spin-fluctuation pairing}

At this point, we do not want to give a comprehensive discussion of the properties of the Hubbard model\cite{Hubbard1963} from a theoretical perspective. This is subject to recent research and currently, there is no full consensus of the correct ground state of this model since it is not solvable exactly\cite{LeBlanc2015}.
How magnetism occurs from this model is well studied by mean-field approaches and even working out the exact solution of the two-site Hubbard model can give important insights, see Chapter 3 of Ref.~\cite{Pavarini2024}. Here, we just want to sketch the way Cooper pairing can be obtained within a weak-coupling approach.

\begin{figure}[tb]
\centering
 \includegraphics[width=\linewidth]{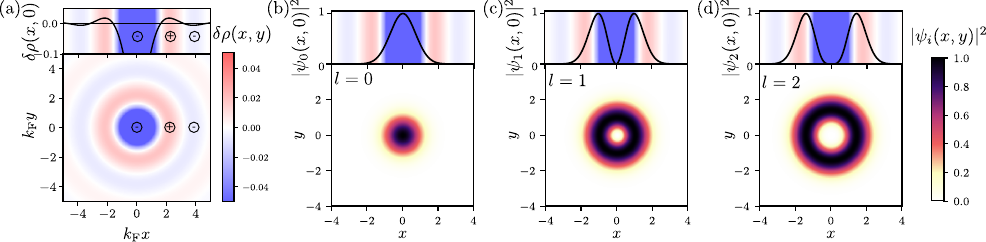}
\caption{Pairing mechanism from bare repulsive interactions: (a) Friedel oscillations in a (2D) metal: A charge in a Fermi liquid produces density modulations which oscillate with period $2k_\mathrm F$ and exhibit sign changes (red/blue). (b-d) Wavefunctions $\psi_l(x,y)$ to angular momentum $l$ in two dimensions demonstrating that $l>0$ states can take advantage of the positive charge modulation $\delta\rho(x,y)$.}
\label{fig_Kohn_Lutt}
\end{figure}

Pairing from repulsive interaction was already studied by Kohn and Luttinger\cite{Kohn1965} who considered a Fermi liquid with repulsive interactions. The key conceptual observation is that a charge in a metal polarizes the gas of free electrons and generates charge density oscillations. While the total charge is compensated due to the atomic cores, there is a charge modulation $\delta\rho(x,y)$ with positive and negative regions. An electron that is in the correct orbit can make use of the attractive part of the charge modulation to lower its energy.
Looking at Fig.~\ref{fig_Kohn_Lutt}, one sees that the $l=0$ state has maximal density at $(x,y)=(0,0)$ and will exhibit repulsive interaction from the huge negative value of $\delta\rho(0,0)$. For higher angular momentum, the electron wavefunction $\psi_l(x,y)$ is suppressed at $(x,y)=(0,0)$ and might exhibit maxima where the density modulations $\delta\rho(x,y)$ are positive: This creates an effective attractive interaction for $l>0$ channels and it is feasible that bound states exist.

Let us try to follow the basic argumentation on the example of the Hubbard model here. Indeed, the bare interaction as in Eq.~(\ref{eq_Hubbard}) yields a pairing interaction that is momentum-independent and repulsive, i.e. $V(\kv,\kv')=U$ as we have already seen it in Eq.~(\ref{eq_tJ}) i.e. no superconductivity emerges. 
Considering now how the free electron gas responds to external perturbations, one can calculate the susceptibility with the result
\begin{equation}
\chi_0(\qv,\omega)=-\sum_\kv \frac{f(\epsilon_{\kv+\qv})-f(\epsilon_\kv)}{\epsilon_{\kv+\qv}-\epsilon_\kv+\omega+i\eta}\,.
\label{eq_bare_susc}
\end{equation}
This quantity describes how the electron gas responds to a perturbation at a given momentum $\qv$ with frequency $\omega$ and takes into account all (virtual) processes of an electron at $\kv+\qv$ with energy 
$\epsilon_{\kv+\qv}$ to scatter into a state at $\kv$ with energy $\epsilon_\kv$. The Fermi functions $f(\epsilon)$ control whether a given process is allowed from the Pauli principle, i.e. whether initial state is occupied and final state is unoccupied and the denominator has a real and imaginary part. The imaginary part becomes a $\delta$ function in the limit of $\eta\rightarrow 0$ and expresses Fermi's golden rule for actual excitations, while the real part gives rise to an energy renormalization due to virtual processes (fluctuations).
The latter is important to generate the effective interaction such that from now on we only consider the static limit $\omega=0$ where the imaginary part vanishes but the real part reaches a finite value. Before getting to the details of the effective interaction and pairing, we want to understand the properties of the susceptibility: States close to the Fermi level contribute most because when choosing the initial state just below $\epsilon_{\kv+\qv}<0$ and the final state just above the Fermi level, $\epsilon_{\kv}>0$, the numerator can reach the value $1$ for low temperatures, while the denominator, i.e. the energy differences can be small. So in order to perform the integral, one basically has to consider all pairs of scatterings between points on the Fermi surface.

\begin{figure}[tb]
\centering
 \includegraphics[width=0.5\linewidth]{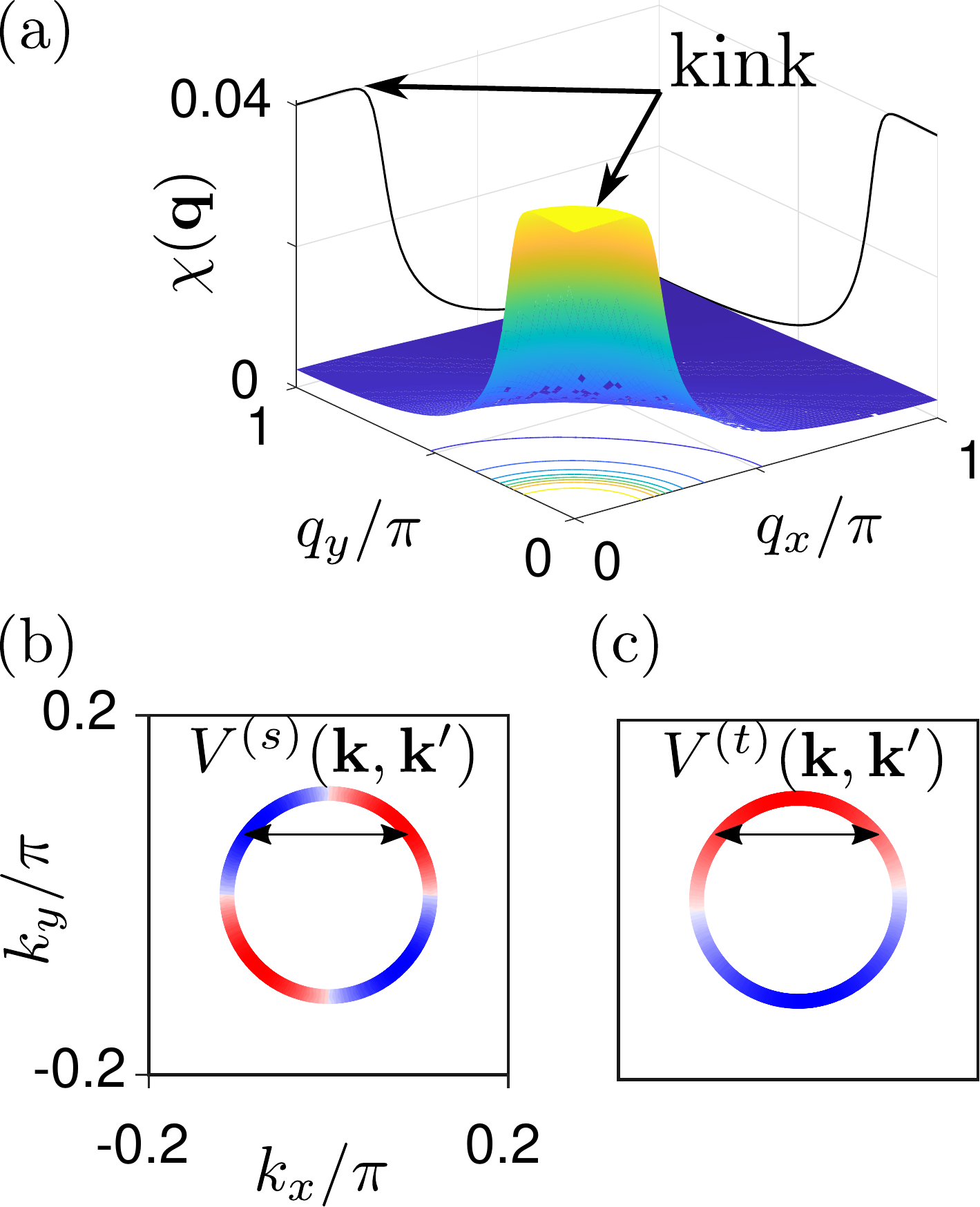}
\caption{Spin-fluctuation pairing at low filling: (a) The susceptibility $\chi(\qv)$ exhibits a plateau with a small maximum at $|\qv|=2 k_{\mathrm F}$ that can drive both singlet and triplet superconducting states, depending on the lattice effects beyond the parabolic dispersion. 
(b) The singlet pairing interaction is repulsive for all pairs $\kv$ and $\kv'$, so it drives sign changes from the maximum of $\chi(\qv)$. (c) This is opposite to the triplet pairing channel that is attractive, but the parity condition requires a sign change.}
\label{Kohn_Luttenger_pairing}
\end{figure}

For a parabolic dispersion $\epsilon_\kv=\frac{k^2}{2m}-\mu$, there is a continuum of processes on a disk of radius $|\qv|=2k_\mathrm F$ with a cusp or kink towards $2k_\mathrm F$, see Fig.~\ref{Kohn_Luttenger_pairing}(a).

The key step is that the effective interaction gets modified in perturbation theory to \cite{Scalapino2012,Romer2015}
\begin{equation}
 V(\kv,\kv')=U+U^2\chi_0(\kv-\kv')
\end{equation}
and therefore the coefficients $V_l$ in Eq.~(\ref{expansion_Vkk}) of the angular expansion become nonzero with the result of a finite $T_c$ for pairing in higher angular momentum $l>0$.

\begin{figure}[tb]
\centering
 \includegraphics[width=\linewidth]{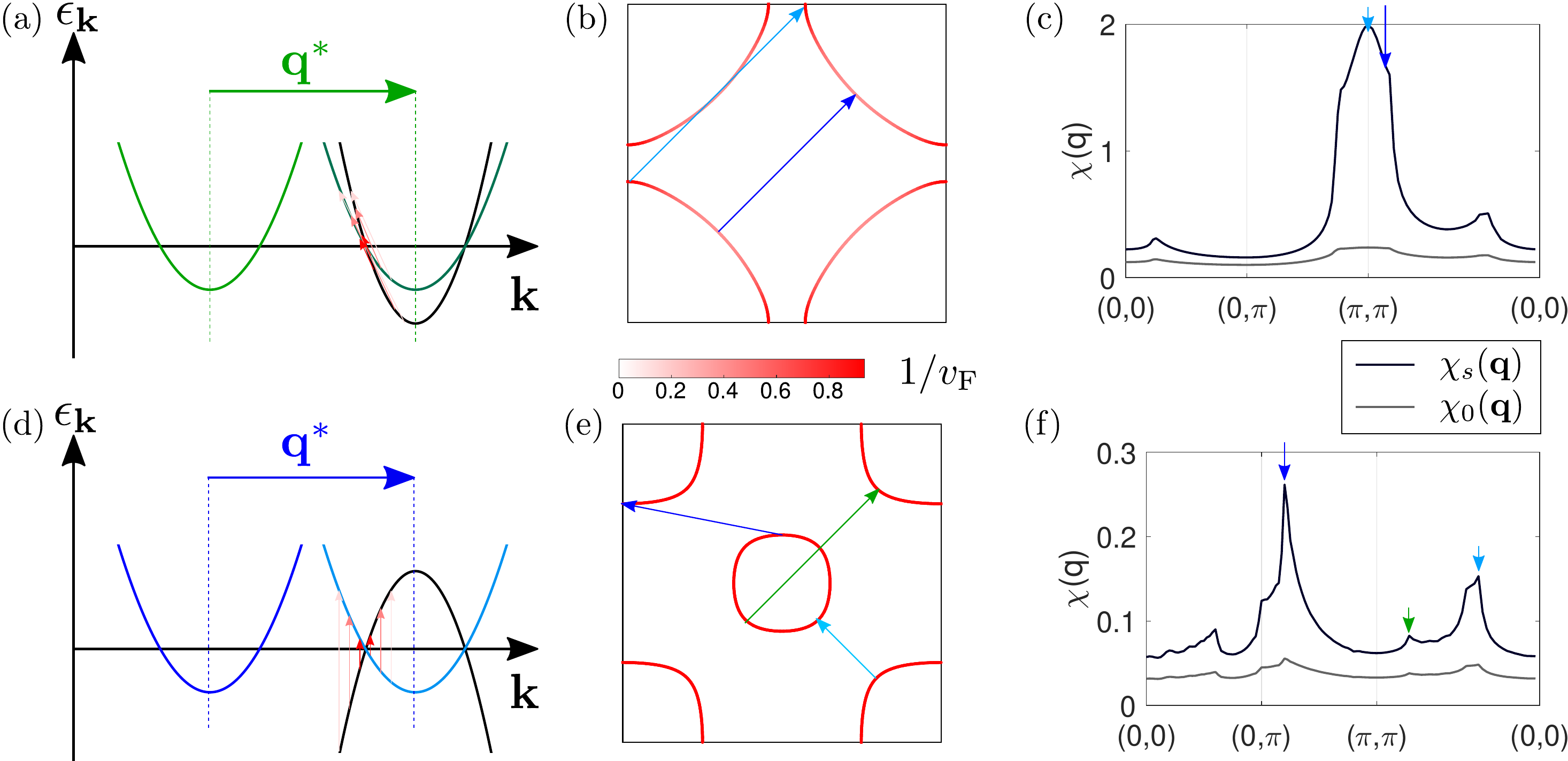}
\caption{Fermi surface nesting: (a) Perfectly overlapping parabolic electron-like bands with nesting vector $\qv^*$ yield contributions to the susceptibility that are not \emph{vertical}, i.e. spread out in a range of q-vectors as symbolized by red arrows that connect filled with empty states. (d) Overlapping electronlike and hole-like bands give vertical scattering contributions from the condition $\epsilon_\kv\approx -\epsilon_{\kv+\qv^*}$ (red) arrows that are weighted according to the energy difference (lighter red arrows) with preference of small $v_\mathrm F$, i.e. flat bands. (b) Example of a single-band model $t'=-0.25t$, (Fig.~\ref{fig_lifshitz}), where nesting vectors connect parts of the Fermi surface with opposite Fermi velocity (blue arrows) and yield large contributions in the susceptibility (c). (e) When two electronlike pockets are present, the nesting with good overlay (green arrow) gives only small contributions, while processes with opposite Fermi velocity are dominant (blue arrows).}
\label{FS_nesting}
\end{figure}

For electrons on a lattice, the geometry of the Fermi surface then condenses important processes to a single momentum transfer $\qv$ where $\chi(\qv) $ becomes enhanced more. This happens because of shape, i.e. the Fermi surface has more flat parts, or because of different Fermi surface sheets that when displaced by a fixed vector $\qv^*$ have overlapping shapes. This scenario yields many scattering processes at the single momentum transfer, but is only enhanced if the Fermi velocities of segments are opposite as graphically shown in Fig.~\ref{FS_nesting}(a,d) and numerically observed for two choices of example band structures.

The same happens if there are saddle points in the dispersion close to the Fermi level where many more states with similar $\kv$ are within the range of energy relevant for virtual particle-hole excitations or if there are multiple Fermi surfaces (with similar shape) that are displaced with a finite $\qv$ (Fermi surface nesting).

If the bare interaction $U$ is sizable, second order perturbation theory might not be sufficient any more; a way out is the random phase approximation where a selection of processes are summed up to all orders and rearranged into a geometric series,
\begin{equation}
 \chi_\mathrm s(\qv)=\frac{\chi_0(\qv)}{1-U\chi_0(\qv)}.\label{eq_susc_RPA}
\end{equation}
For those momenta $\qv$ where the susceptibility is already large, the multiple interaction processes further enhance the response, see Fig.~\ref{FS_nesting}(c,f) until it diverges in the limit that the denominator vanishes, $U\chi_0(\qv) \rightarrow 1$. In this regime, the perturbative approach is not justified any more, but analyzing the structure of $\chi_\mathrm s(\qv)$, how it relates to the energy dispersion $\epsilon_\kv$ and symmetries of the system is still meaningful.

The effective interaction that enters Eq.~(\ref{eq:BCS}) is then given by \cite{Scalapino2012,Romer2015}
\begin{equation}
 V(\kv,\kv')=U+\frac 12 \left [3 U^2\chi_\mathrm s(\qv)-U^2\chi_\mathrm c(\qv)\right]\label{eq_sf_pairing_vertex}
\end{equation}
where now $\chi_\mathrm c(\qv)=\frac{\chi_0(\qv)}{1+U\chi_0(\qv)}$ is the resummed susceptibility for density perturbations (charge susceptibility). Without going into details of the origin of the two terms in the effective interaction, let us mention that the charge susceptibility does not exhibit a divergence as a function of $U$ and its effect on the superconducting order parameter is therefore weak within our framework.
When performing the projection into singlet and triplet interactions, Eq.~(\ref{eq_pairing_symm}), one sees that pairing between electrons $\kv-\kv'=\qv$ has the functional form of $V^s(\qv)\sim U + \frac 32 U^2\chi_s(\qv)$ for singlet pairing and $V^t(\qv)\sim -\frac 12 U^2\chi_s(\qv)$. The first property means that the interaction is positive (repulsive) for all combinations of $\kv$ and $\kv'$, so the order parameter becomes large and of opposite sign for $\kv$ and $\kv'$ where the susceptibility has a maximum, see Fig.~\ref{Kohn_Luttenger_pairing} (b). In the triplet case, the interaction is negative (attractive), has a smaller prefactor, i.e. pairing instabilities from spin fluctuations are usually weaker and only present in certain regions of the phase diagram\cite{Romer2015}. The maxima of the order parameter are connected by maxima in the susceptibility, but have the same sign. For triplet order parameters, a sign change is however imposed by the Pauli principle which makes the order parameter odd in momentum $\Delta^t_\kv=-\Delta^t_{-\kv}$, see Fig.~\ref{Kohn_Luttenger_pairing} (c).

In summary, we can state that not only the electronic structure picks out a leading pairing state if the pairing interaction is kept constant, but also the pairing interaction from electron-electron interactions itself determines the effective pairing glue and its momentum dependence.

\subsubsection{Multiband pairing}
We conclude this section with a few remarks towards current research questions, additions for more realistic descriptions, but refer to review material showing example results\cite{Scalapino2012,HirschfeldCRAS,Pavarini2024}. Real materials often possess extra electronic degrees of freedom (with quantum number $\ell_i$), such as orbital, sublattice, valley, resulting in multiband electronic structures. In such systems, the pairing interaction acquires a matrix structure, i.e. the effective pairing Hamiltonian reads
\begin{equation}
 H_{\text{int}}=\frac{1}{2}\sum_{\kv,\kv'}
 V_{\ell_1\ell_2\ell_3\ell_4}(\kv,\kv')
 c_{\kv' \ell_1\up}^\dagger c_{-\kv' \ell_2 \down}^\dagger
 c_{-\kv \ell_3 \down} c_{\kv \ell_4 \up} 
\end{equation}
where the pairing interaction from spin fluctuations is given by
\begin{equation}
 	V_{\ell_1\ell_2\ell_3\ell_4} (\kv,\kv')\! =\!\frac 12 \left[3\bar U^\mathrm s \bar\chi_\mathrm s (\kv-\kv') \bar U^\mathrm s 
 	+   \bar U^\mathrm s - \bar U^\mathrm c \bar\chi_\mathrm c (\kv-\kv') \bar U^\mathrm c +  \bar U^\mathrm c \right]_{\ell_1\ell_2\ell_3\ell_4}\label{eq_Gamma_l}
\end{equation}
Here, all quantities with bar are matrices that describe the generalization of the Hubbard interaction $\bar U_c$ and $\bar U_s$ and the charge and spin susceptibility, but resolved by the additional degrees of freedom (orbitals), $\bar\chi_{c,s}(\qv)$. Each electronic degree of freedom has a different interaction strength and different tendency to form Cooper pairs\cite{Mazin2009,Scalapino2012}. This allows for a much richer variety of superconducting states,  including interband pairing giving nontrivial opening of gaps at finite energy, see Fig.~\ref{quasiparticle_dispersion}.

The role of electronic correlations has not been discussed here so far. However, in many strongly correlated materials, the weak-coupling picture is not justified any more: Already the normal state itself deviates from conventional Fermi-liquid behavior, making it difficult to formulate the pairing problem in terms of well-defined quasiparticles and a weak residual interaction. In these cases, understanding superconductivity requires approaches that go beyond the Fermi-liquid paradigm.

\subsection{Other pairing mechanisms}

In the previous section, we have explained the emergence of the superconducting order parameter on the example of spin-fluctuation pairing mechanism, but need to mention some other mechanisms to give a balanced overview of current research. Often, the interplay of multiple mechanisms is discussed, so also possible cooperative interplay with electron-phonon pairing\cite{Nunner1999}. Charge fluctuations are already included in Eq.~(\ref{eq_sf_pairing_vertex}). In a similar picture as spin-fluctuations also the charge fluctuations become large when the system is close to a charge-density-wave instability and can drive superconducting instabilities \cite{Roemer2022}.

Other proposals are pairing mediated by nematic fluctuations, which emerge near electronic nematic quantum critical points\cite{Fradkin2010} where rotational symmetry is spontaneously broken. This has been discussed in conjunction with Fe-based superconductors\cite{Fernandes2014} following up by the idea that quantum critical fluctuations can serve as a source of strong attractive interactions, a recent example are fluctuating loop-current order, a form of spontaneous current in the ground state\cite{Palle2024}.

\section{Experimental probes}

In the previous two sections, we have examined the properties of the superconducting order parameter, the consequences for the quasiparticle excitations in a superconductor and explained some mechanisms that can drive such unconventional Cooper pairing. Now, we want to follow up with the question how $\Delta_\kv$ can be detected experimentally and explain why it is so difficult to pinpoint the superconducting order parameter and even more subtle to find evidence for the pairing mechanism. The pairing interaction cannot be measured directly to find out which is the pairing glue, so we only summarize some techniques to detect the superconducting order parameter $\Delta_\kv$.

\subsection{Detection of parity}

The first question to ask is whether the pairing is spin singlet or spin triplet. In nuclear magnetic resonance experiments\cite{Gosar2024}, the precession frequency of a nuclear spin in an external field is measured. This frequency is shifted from the one of a free spin by the shielding of the surrounding electrons. Once the electrons pair up to singlet states in a superconductor, the electron spins do not respond to the field any more, and the shielding, or the Knight shift drops to zero at zero temperature when all electronic states are paired. In a triplet state, the Cooper pairs form a $S=1$ state which still shields the nuclear spin in some way, although there are some subtleties regarding the direction of the external field relative to the spin of the Cooper pairs\cite{Sigrist_notes}.

\subsection{Phase insensitive probes}

A number of experimental techniques probe the quasiparticle dispersion, Eq.~(\ref{eq_qp}) and are therefore insensitive to the phase of the order parameter because it only depends on the amplitude $|\Delta_\kv^{s/t}|$. Among those are tunneling experiments where the low-energy density of states, Eq.~(\ref{eq_dos_SC}) is measured and features of the order parameter as gap minima or gap maxima can be read off, see Fig.~\ref{nontrivial_gaps_single_band}. Angle-resolved photoemission experiments measure the electron density of states resolved in momentum space, so can map out $E_\kv$ in the Brillouin zone and from comparison to the normal state then deduce $|\Delta_\kv|$. 

Thermodynamic measurements are usually insensitive to the sign of the order parameter as well. This can be seen on the example of the specific heat which is the change of the entropy \begin{equation}
S=-2k_\mathrm{B}\sum_{\kv}[f(E_\kv)\ln f(E_\kv)+(1- f(E_\kv))\ln (1-f(E_\kv))],                                                                                                                                           \end{equation}
and therefore inherently only depends on $E_\kv$. The specific heat $C(T)=T dS/dT$ then probes the density of states, but is only sensitive to the magnitude of the order parameter as well. For a fully gapped superconductor, one obtains exponential temperature dependence and for nodal superconductors a power law dependence; discriminating between them can become challenging experimentally because low enough temperatures need to be reached. In any case, one can only constrain the order parameters roughly to the two classes fully gapped and nodal, but not detect the sign or symmetry of it. For example, a sign-changing $s$-wave order parameter can be fully gapped or exhibit accidental nodes, even a smooth transition between these states is possible. The same is true for nontrivial superconducting states when different Fermi surface topologies are allowed: A circular Fermi surface centered at the $\Gamma$ point has to exhibit nodes, but Fermi surfaces at the $X$ and $Y$ point can lead to a fully gapped system if the nodes of the order parameter are not hitting the Fermi surface, see Fig.~\ref{fig_trivial_irrep}. Other measurements of this type are low-temperature penetration depth $\lambda$ and thermal conductivity giving power law dependence for nodal states \cite{Sigrist_notes}.

\subsection{Phase sensitive probes}

Direct phase sensitivity can be achieved by bringing phase coherent superconductors to interfere. This can happen in inhomogeneous, granular superconductors where islands of superconductivity form, if the phase of the order parameter changes sign, there are associated spontaneous currents that produce local magnetization. Such effects are only expected in nontrivial superconductors, a clear signature is the \emph{paramagnetic} response of the superconductor at low fields rather than diamagnetic, i.e. the magnetic susceptibility increases in the superconducting state instead of becoming negative. This Wohlleben effect has been observed in unconventional superconductors \cite{Braunisch1992}, other associated phenomena are under discussion in recent research \cite{Zi-Xiang2021,Andersen2024}.

More controlled experimental verifications use the symmetry of the order parameter and the fact that in nontrivial superconductors, the order parameter breaks a lattice symmetry: Fabricating tunneling junctions between a nontrivial superconductor and a known conventional ($s$-wave) superconductor, one can construct situations where the overlap between the superconducting order parameters vanishes by symmetry, so no Cooper pair tunneling is allowed similar to the case of vanishing hopping elements of tight-binding Hamiltonians, see Fig.~\ref{fig_overlap}. 
Such corner Josephson junction experiments\cite{Wollman1993}
are one of the most direct phase-sensitive demonstrations of $d$-wave pairing in the cuprates and variants of these are used to exclude nontrivial pairing symmetries\cite{Zhang2009} or examine pairing symmetries by rotating (twisting) $d$-wave superconductors\cite{Can2021}.

\subsection{Indirect phase sensitivity}

Probing the response to magnetic perturbations is a common way to learn about the electronic structure and especially how it changes when entering the superconducting state. Indeed, when calculating the spin susceptibility of a superconductor, not only the eigenvalues $E_\kv$, but also the eigenstates as given in Eq.~(\ref{eq_bogo_trafo}) enter. Here we give a well known result for a single band superconductor with singlet order parameter $\Delta_\kv$\cite{Dai2024},
\begin{align}
   \chi^{+-}_0(\mathbf{q},\omega)  =
    \sum_{\mathbf{k}, E>0}&\left[ \left(1  -  \frac{\epsilon_{\mathbf{k}}\epsilon_{\mathbf{k}  +  \mathbf{q}}  +  \Delta^*_{\mathbf{k}  +  \mathbf{q}}\Delta_{\mathbf{k}}}{E_{\mathbf{k}}E_{\mathbf{k}  +  \mathbf{q}}}\right)   \frac{1  -  f(E_{\mathbf{k}})  -  f(E_{\mathbf{k}  +  \mathbf{q}})}{\omega  +  E_{\mathbf{k}  +  \mathbf{q}}  +  E_{\mathbf{k}}  +  i\eta}\right.
     \notag\\&
      +  \left(1  -  \frac{\epsilon_{\mathbf{k}}\epsilon_{\mathbf{k}  +  \mathbf{q}}  +  \Delta^*_{\mathbf{k}  +  \mathbf{q}}\Delta_{\mathbf{k}}}{E_{\mathbf{k}}E_{\mathbf{k}  +  \mathbf{q}}}\right)  \frac{f(E_{\mathbf{k}})  +  f(E_{\mathbf{k}  +  \mathbf{q}})  -  1}{\omega  -  E_{\mathbf{k}  +  \mathbf{q}}  -  E_{\mathbf{k}}  +  i\eta}
     \notag\\
    &   +     \left(1 + \frac{\epsilon_{\mathbf{k}}\epsilon_{\mathbf{k}  +  \mathbf{q}} + \Delta^*_{\mathbf{k}  +  \mathbf{q}}\Delta_{\mathbf{k}}}{E_{\mathbf{k}}E_{\mathbf{k}  +  \mathbf{q}}}\right)\frac{f(E_{\mathbf{k}}) - f(E_{\mathbf{k}  +  \mathbf{q}})}{\omega + E_{\mathbf{k}  +  \mathbf{q}} - E_{\mathbf{k}} + i\eta}
    \notag\\
    & +   \left. \left(1 + \frac{\epsilon_{\mathbf{k}}\epsilon_{\mathbf{k}  +  \mathbf{q}} + \Delta^*_{\mathbf{k}  +  \mathbf{q}}\Delta_{\mathbf{k}}}{E_{\mathbf{k}}E_{\mathbf{k}  +  \mathbf{q}}}\right)\frac{f(E_{\mathbf{k}  +  \mathbf{q}}) - f(E_{\mathbf k})}{\omega + E_{\mathbf{k}} - E_{\mathbf{k}  +  \mathbf{q}} + i\eta}\right].
    \label{eq_2_1}
\end{align}
This quantity is basically describing how the superconductor reacts to a perturbation at a given frequency $\omega$ with momentum $\qv$, its imaginary part tells how energy can be absorbed by the system. Looking at the terms, it turns out that these can be divided into two contributions: The first two lines require $\omega\approx E_{\kv+\qv} + E_{\kv}$ to produce a $\delta$-function contribution (Fermi's golden rule), while the last two lines contribute for $\omega \approx 0$ because of the different sign in the energy denominator.

\begin{figure}[tb]
\centering
 \includegraphics[width=0.5\linewidth]{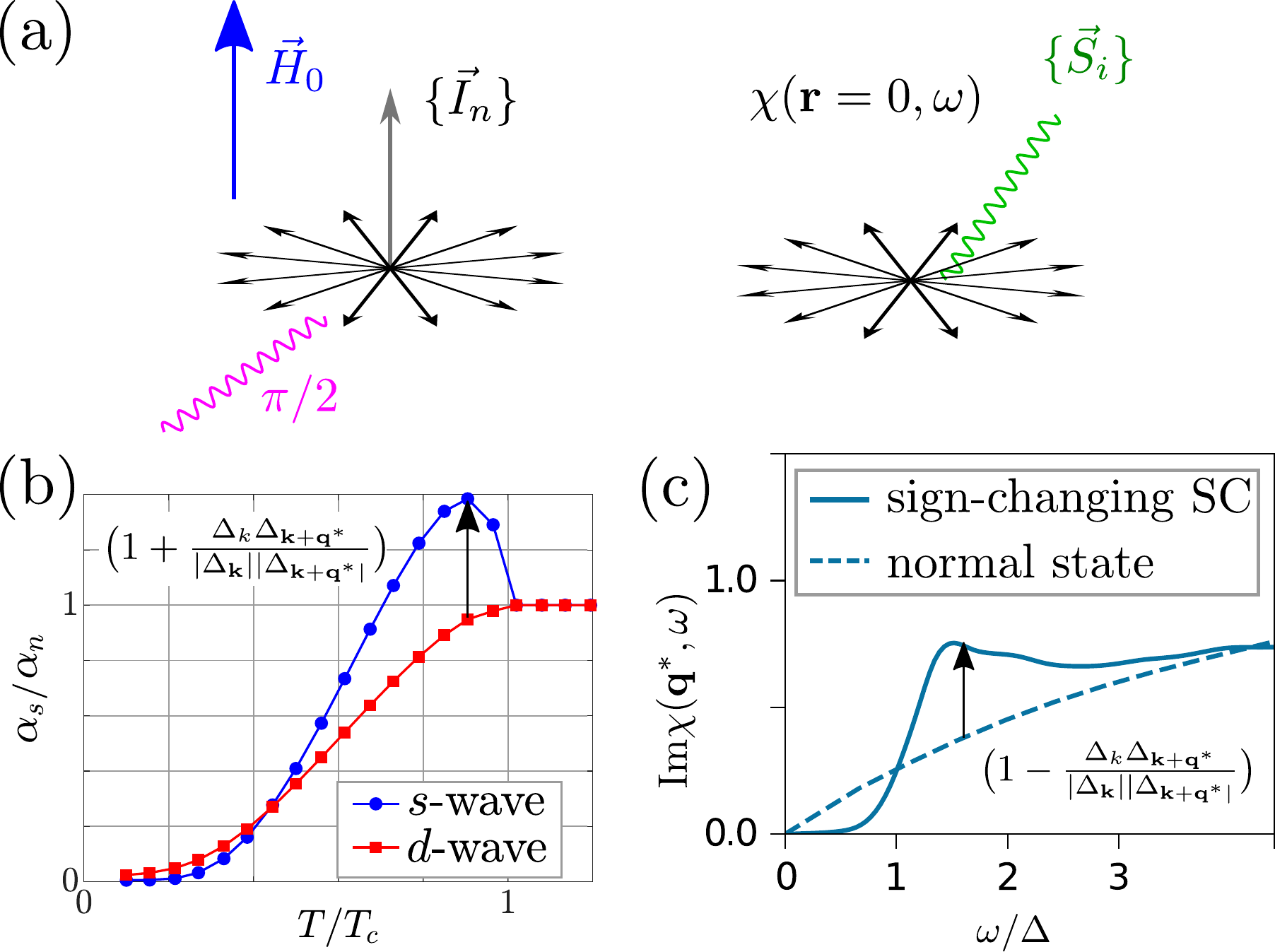}
\caption{(a) Schematics of a nuclear spin relaxation experiment: Nuclear spins $\mathbf I_n$ are subject to an external magnetic field $\mathbf H_0$ and align (gray arrow). A microwave pulse of tuned length rotates the spins by 90$^\circ$, so they precess in the plane perpendicular to the field (black arrows). The interaction with the electron spins $\mathbf S_i$ allows one to probe the spin susceptibility of the electronic structure. (b) Relaxation rate, i.e. decay rate back to the ground state as calculated for two different superconducting pairing states: In the sign-changing $d$-wave case, the relaxation rate drops due to the gapping of the spectrum. In the $s$-wave case without sign change, the coherence factor enhances the relaxation rate before it eventually drops at low temperatures. (c) Neutron resonance in a sign-changing superconductor: The susceptibility at finite energy $\omega$ is enhanced at energies $2\Delta$ due to the coherence factor. At low energies, eventually the gapping of the states reduces the susceptibility compared to the normal state\cite{Chen2019}.}
\label{phase_sensitive}
\end{figure}

\paragraph*{Type I coherence factors}
Similar to the normal state response, Eq.~(\ref{eq_bare_susc}), the largest contribution comes from the states close to the Fermi level, where we can replace $E_{\mathbf{k}}\rightarrow |\Delta_{\mathbf{k}}|$ and see that the contributions from superconductivity have the form
\begin{equation}
 \left(1  - \frac{\Delta_{\mathbf{k}  +  \mathbf{q}}\Delta_{\mathbf{k}}}{|\Delta_{\mathbf{k}}||\Delta_{\mathbf{k}  +  \mathbf{q}}|}\right)\approx\Bigl\{\begin{array}{l}2 \quad \text{different sign} \\
 0\quad  \text{same sign}                                                                                                                                         \end{array}
\end{equation}
for the first two lines. The spin response at energy $\omega$ and momentum transfer $\qv$ can be measured in neutron scattering experiments where an additional resonance peak at $\omega$ below $\Delta_{\text{max}}$ is interpreted as a hallmark of a sign-changing order parameter, see Fig.~\ref{phase_sensitive} (c).

\paragraph*{Type II coherence factors}
The second two lines in our expression have the opposite behavior,\vspace{-0.4cm}
\begin{equation}
 \left(1  + \frac{\Delta_{\mathbf{k}  +  \mathbf{q}}\Delta_{\mathbf{k}}}{|\Delta_{\mathbf{k}}||\Delta_{\mathbf{k}  +  \mathbf{q}}|}\right)\approx\Bigl\{\begin{array}{l}                                                                                                            2 \quad\text{same sign}\\
 0\quad\text{different sign}                                                                                                                         \end{array}
\end{equation}
leading to an enhancement of the response just for the case of a conventional superconductor without sign change. In a spin-relaxation experiment of nuclear spins, one prepares the spins to be anti-aligned with the external field by a microwave pulse, see Fig.~\ref{phase_sensitive} (a). The nuclear spins then interact with the electronic degrees of freedom and relax back to their ground state. Once the system enters the superconducting state, the density of states abruptly changes at low energies, a tiny enhancement for compensated order parameters can occur. However, the type II coherence factors yield a strongly enhanced relaxation of the nuclear spins to the ground state (aligned with the external field) as indication for non-sign-changing superconductivity, see Fig.~\ref{phase_sensitive} (b), although a number of details need to be considered \cite{Dai2024,Gosar2024}.

\section{Conclusions}

In this review article, we have used concepts to describe the electronic structure of unconventional superconductors  in terms of tight-binding models, introduced the statistics of fermions that play an essential role when using the microscopic approach for interacting systems. The gap of superconducting systems is connected to the anomalous expectation values which we calculated within some model systems to understand the difference between conventional and unconventional superconductors.
Guided by symmetries, we have discussed nontrivial superconducting pairing states and examined the quasiparticle excitation spectra in more detail. A key observation is that the excitation energies only depend on the magnitude of the order parameter. To see consequences of unconventional pairing, one needs to examine the eigenstates that describe how electrons and holes are superimposed for quasiparticle excitations. Finally, we walked through some experimental probes and explained why and how these can probe the nature of the superconducting state.

Research on unconventional superconductivity spans many decades, is triggered by new discoveries of materials and developments on the experimental side. Often, there are twists in the interpretation of data \cite{Maeno2024} and the identification of the superconducting order parameter, not to mention the difficulties in pinpointing the pairing mechanism. We hope to have given some guidance what can be said about unconventional superconductors from experimental data and which parts are subtle and require further investigation. This was done by reviewing some core principles for the formation of the superconducting state that should allow the reader to access some more specialized literature on this topic and understand recent developments in the field.


%
%
\section*{Acknowledgments}
I thank Peter J. Hirschfeld and Brian M. Andersen for discussions and comments on the manuscript.

\section*{Disclosure statement}
No potential conflict of interest was reported by the author(s).

\section*{Notes on contributor}
Andreas Kreisel is an Associate Professor at Uppsala University, Sweden. He completed his studies of physics at the Goethe University in Frankfurt with the degree of diploma and subsequently received his PhD (Dr.~phil.~nat.) from the same university. Afterward, he performed postdoctoral research at the University of Florida and at the Niels Bohr Institute in Copenhagen. Then he moved to the University of Leipzig where he completed the Habilitation in 2020. After a position as assistant professor at the Niels Bohr Institute, he moved to Uppsala University in 2026. His research is centered around the understanding of the interacting electron problem with applications to quantum magnetism and unconventional superconductivity to understand properties of real materials.
%
%
%
%
%
%
%

%

\bibliographystyle{phd}
\bibliography{references_AK}

\section{Appendices}

\label{sec_app}
\subsection{Electronic states in solids}
\label{electronic_states}

\begin{figure}
\centering
\includegraphics[width=0.6\linewidth]{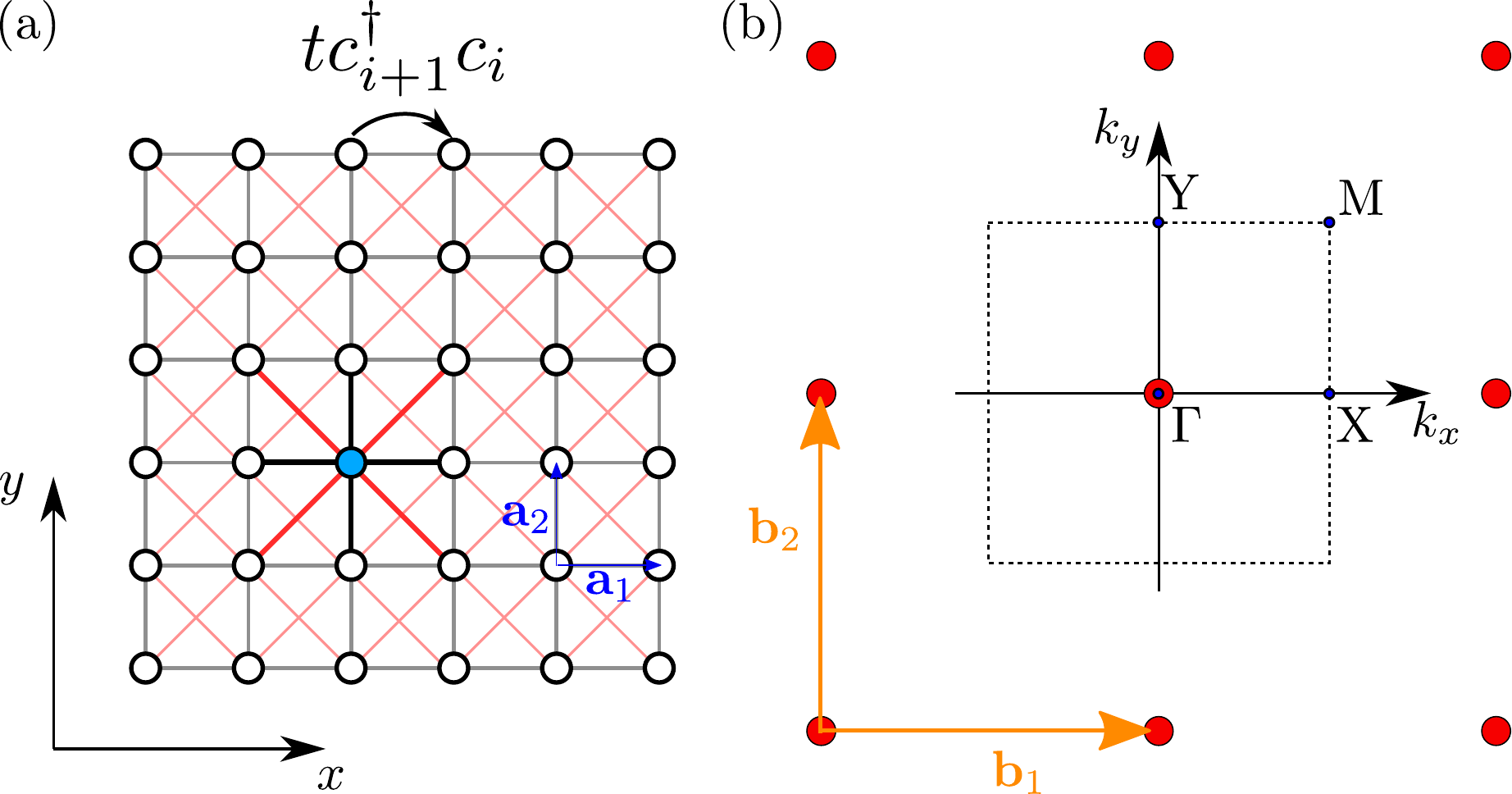}
\caption{(a) Model of electrons on a square lattice with the lattice points (open circles) which are translationally invariant with the primitive vectors $\mathbf a_i$. We consider hoppings of electrons to the nearest neighbors (gray lines) and next nearest neighbors (rose lines). The thick lines connecting the lattice point marked blue are representing the $4+4=8$ hopping processes from and to that lattice point. (b) Corresponding reciprocal lattice (red points) that are connected by the reciprocal lattice vectors $\mathbf b_i$. One can describe all Bloch states using the crystal momenta within the Brillouin zone (dashed line). For our example system that has tetragonal symmetry, the high symmetry points $\Gamma$, X, M, Y are marked with small blue dots.} \label{fig_real_reciprocal}
\end{figure}

To set the stage for understanding superconductivity and how the energy gap in unconventional superconductors arises, we need to introduce some concepts. This does not aim to be self-contained, so the reader is referred to  textbooks on solid state physics \cite{Kittel2004,Girvin_Yang_2019, Coleman2015} or lecture notes for further reading \cite{timm2020superconductivity, Sigrist_notes}.

In the approximation of non-interacting particles, the electronic states in crystals are described by the time-independent Schrödinger equation,
$H\psi(\mathbf{r})=E\psi(\mathbf {r})$, 
where the Hamiltonian is the sum of the kinetic energy and the periodic potential $ V(\mathbf {r}+\mathbf{R})=V(\mathbf{r})$. The lattice vectors are given in terms of the $d$ primitive lattice vectors $\mathbf a_i$ in a $d$ dimensional crystal, i.e. we can write $\mathbf R=\sum_i n_i \mathbf a_i$ where $n_i$ are integers.

To make use of the lattice periodicity, we introduce the reciprocal lattice with lattice vectors $\mathbf G=\sum_i m_i \mathbf b_i$ where $\mathbf b_i$ are the primitive reciprocal lattice vectors. These are defined by satisfying the orthogonality relation $\mathbf a_i\cdot \mathbf b_j=2\pi \delta_{ij}$.

As first shown by Felix Bloch, it is possible to write the eigenstates of band $n$ of a periodic crystal as $\psi_{n\mathbf{k}}(\mathbf{r}) = e^{i\mathbf{k}\cdot\mathbf{r}}u_{n\mathbf{k}}(\mathbf{r}),$
where the function $u_{n\mathbf{k}}(\mathbf{r})$ has the periodicity of the lattice. The crystal momentum $\mathbf k$ plays the role of the wavevector of a free particle in analogy to the exponential factor of a plane wave. It has the properties that it is periodic with the introduced reciprocal lattice vectors because of the periodicity of the Bloch wavefunction $\psi_{n\mathbf{k}}(\mathbf{r})=\psi_{n\mathbf{k}+\mathbf{G}}(\mathbf{r})$ and quantized due to the confinement in the (large) but finite solid.

All allowed crystal momenta $\mathbf k$ are within a Brillouin zone, see Fig.~\ref{fig_real_reciprocal} (b) for a simple example. Often one refers to the first Brillouin zone, which contains all momenta that are closest to the $\Gamma$ point at $\mathbf k=0$. However, for practical applications, it is convenient to use a parallelepiped spanned by the reciprocal lattice vectors, i.e. the crystal momentum can be parametrized by $\mathbf k=\sum_i h_i \mathbf b_i$ where $h_i \in [0\cdots 1)$, i.e. the area of relevant crystal momenta is shifted (and potentially rearranged in shape).

\begin{figure}
\centering
\includegraphics[width=\linewidth]{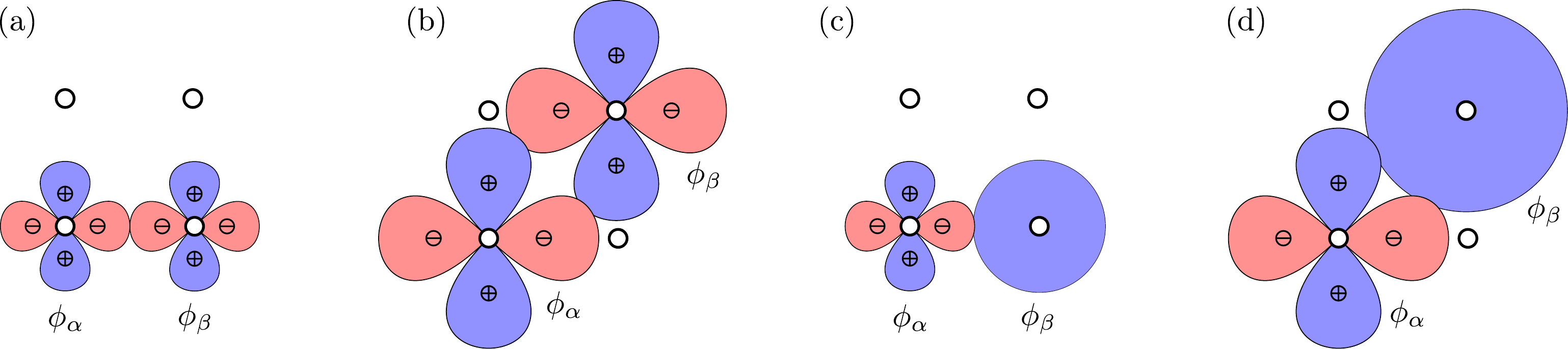}
\caption{Symmetry properties of the hopping matrix element $H_{\alpha\beta}(\mathbf{R})$ for nearest neighbors (nn) and next nearest neighbors (nnn). For orbitals of symmetry $d_{x^2-y^2}$, the wavefunction exhibits positive and negative phases (blue/red). In consequence the nn hopping elements (a) are expected to be of opposite sign than the nnn elements (b) because the relative sign of the overlapping parts of the wavefunctions differ, same sign vs. opposite sign. For overlap of orbitals of different symmetry, some elements are expected to vanish identically as visualized in panel (d).} \label{fig_overlap}
\end{figure}

While the result of Bloch's theorem fixes the general form of the wavefunctions, we need to solve the Schrödinger equation to obtain the energy $\epsilon_n(\mathbf{k})$ of band $n$. In the following, we use the tight-binding approximation to represent the low-energy bands in materials. Here, one  assumes that the electronic states remain strongly localized around atomic sites and can therefore be described using a basis of orbitals $\phi_\alpha(\mathbf{r}-\mathbf{R})$, where $\mathbf{R}$ labels the elementary cell, and $\alpha$ denotes a degree of freedom within the elementary cell, i.e. a placeholder for the orbital type or spin. The Bloch states are lattice-periodic superposition of these localized orbitals,
$\frac{1}{\sqrt{N}}
\sum_{\mathbf{R}}
e^{i\mathbf{k}\cdot\mathbf{R}}
|\phi_\alpha(\mathbf{R})\rangle$, 
where $N$ is total number of unit cells in the crystal. We can then express the Hamiltonian using the matrix elements $ H_{\alpha\beta}(\mathbf{R}-\mathbf{R}')=\langle \phi_\alpha(\mathbf{R}) |H|\phi_\beta(\mathbf{R}') \rangle$.
Translational symmetry implies that these matrix elements depend only on the relative displacement between lattice sites and encode the symmetry of the corresponding wavefunctions as visualized in Fig.\ref{fig_overlap}. The Bloch Hamiltonian is therefore the Fourier transform of the matrix elements,
\begin{equation}
H_{\alpha\beta}(\mathbf k)=\sum_{\mathbf{R}}
e^{i\mathbf{k}\cdot\mathbf{R}} H_{\alpha\beta}(\mathbf{R}),
\end{equation}
and the energy bands follow from solving the eigenvalue problem
$H_{\alpha\beta}(\mathbf k) u_n(\mathbf{k})= \epsilon_n(\mathbf{k})u_n(\mathbf{k})$ for the band $n$.

In practice, the matrix elements are often separated into on-site energies and hopping amplitudes to neighboring lattice points. For a single orbital per site with nearest-neighbour hopping
$t=H(\mathbf{R_{\mathrm{nn}}})$ when $\mathbf R_{\mathrm{nn}}$ is one of the 4 vectors connecting  the nearest neighbours and  the
hopping $t'=H(\mathbf{R_{\mathrm{nnn}}})$ when  $\mathbf R_{\mathrm{nnn}}$ is one of the 4 vectors connecting  the next nearest neighbours.
Summing over the 8 possible processes that connect to the blue lattice point in Fig. \ref{fig_real_reciprocal}(a), one can verify that the Hamiltonian takes the form
$H(\mathbf k)=-2t(\cos k_x +\cos k_y)-4t'\cos k_x \cos k_y$, where we have used the lattice constant $a=1$ as the length unit, and often set $t=1$ to use it as energy unit.

\subsection{Example: Dispersion of a single-band model on the square lattice}
\label{sec_single_band}
\begin{figure}[tb]
\centering
\includegraphics[width=\linewidth]{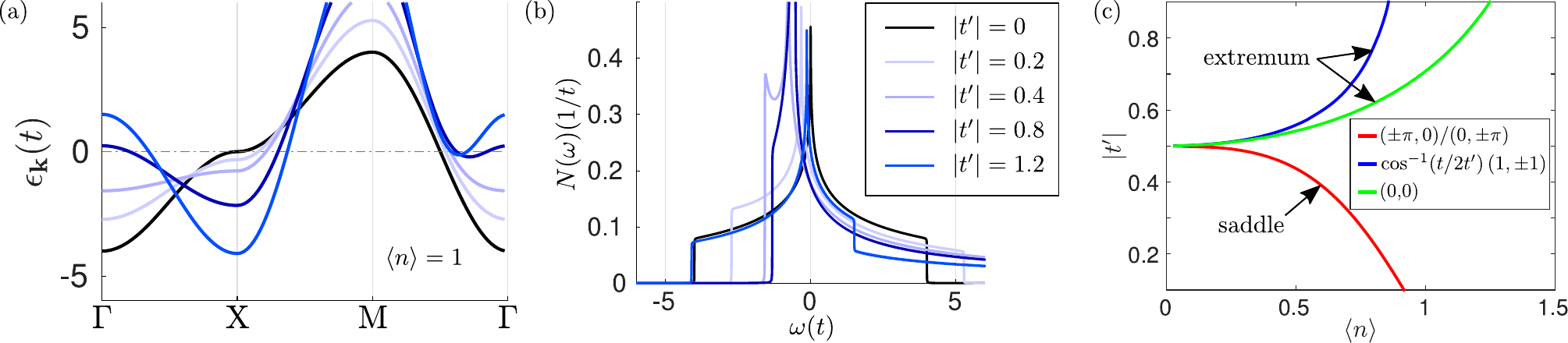}
\caption{(a) Dispersion of the square lattice tight binding model Eq.~(\ref{eq_disp_single}) along a high symmetry path, see Fig.~\ref{fig_real_reciprocal} for different choices of $t'<0$. The chemical potential is tuned to stay at half filling $\langle n\rangle =1$ when summing over spin, $n=n_\uparrow+n_\downarrow$. (b) Corresponding density of states per unit cell $N(\omega)=\frac 1N\sum_\kv \delta(\epsilon_\kv-\omega)$ exhibiting several van Hove singularities from band maxima/minima (steps) and saddle points (peaks). (c) Evolution of the position of band minima and maxima as well as the saddle point as function of filling.}
       \label{single_band_lifshitz}
\end{figure}

The single-band model with nearest neighbor hopping $t$ and next nearest neighbor hopping $t'$ has some relevance in the discussion of high $T_c$ cuprate materials\cite{Scalapino2012}. Starting from the Hamiltonian in real space as introduced in the previous section, see Fig.~\ref{fig_real_reciprocal}, we have obtained the corresponding Bloch Hamiltonian and therefore the band dispersion is 
$ \epsilon_\kv=-2t(\cos k_x +\cos k_y)-4t'\cos k_x \cos k_y -\mu$, as given in Eq.~(\ref{eq_disp_single})
where now $\mu$ is the chemical potential fixing the number of electrons per elementary cell. Allowing various values for $t'$, this dispersion can give rise to several Fermi surface shapes, i.e. the contour lines where $\epsilon_\kv=0$. Fig.~\ref{single_band_lifshitz} shows the band dispersion $\epsilon_\kv$ along high symmetry directions 
for a set of $t'<0$ which is the expected sign for $d_{x^2-y^2}$ orbitals, see Fig.~\ref{fig_overlap}(a,b).
To understand propensity of a given electronic structure towards instabilities such as magnetism or superconductivity, it is instructive to look at the density of states, i.e. the number of states at a given energy which can conveniently be calculated by
\begin{equation}
 N(\omega)=\sum_\kv \delta(\omega-\epsilon_\kv) \label{eq_dos}
\end{equation}
using the Dirac $\delta$ function. For numerical implementations one chooses the Lorentz function representation while for analytical calculations, it is convenient to rewrite it as an integral over one dimension less with the Fermi velocity $|\mathbf{v}_{\mathrm F}(\kv)|=|\mathbf \nabla \epsilon_\kv|$ in the denominator. This comes  from the identity $\delta(f(k))=\sum_i \frac{\delta(k-k_i)}{|f'(k_i)|}$ where $k_i$ are the zeros of $f(k)$ and in the denominator is the derivative $f'(k)$.
Numerical calculations of the density of states of the dispersion in Eq.~(\ref{eq_disp_single}) are shown in Fig.~\ref{single_band_lifshitz}(b).
When a Fermi surface appears, disappears, or becomes disconnected, we observe so-called Lifshitz transitions that are accompanied by nonanalyticities in the density of states.
For small (or large) filling, the dispersion approximates to a parabolic band when expanding around the $\Gamma$ point or ($M$ point), i.e. $\epsilon_{\mathbf k}\approx \frac {{\mathbf k}^2}{2m}-\tilde\mu$ where the effective mass is given by $m=\pm 1/(2[t+2t'])$ and the new chemical potential $\tilde \mu=\mu+4(\pm t+t')$.
The parabolic dispersion in two dimensions is associated with a step in the density of states. One can do similar expansions at a saddle point yielding
$\epsilon_{\mathbf k}\approx a_x\delta k_x^2+a_y\delta k_y^2$ when ignoring the value of the chemical potential and introducing $\delta k_i$ as the distance from the location of the saddle point. Here the second derivatives have opposite signs $\mathrm{sign}(a_x)=-\mathrm{sign}(a_y)$. This yields a logarithmic divergence in the density of states\cite{Hove1953}.

\subsection{Fermion statistics and Pauli principle}
\label{pauli}
So far, we have discussed the electronic state of single electrons in a solid described by the wavefunction $\psi(\mathbf r)$. When describing indistinguishable particles in quantum mechanics, it turns out that the two particle wavefunction $\psi(\mathbf r_1,\mathbf r_2)$ can acquire a phase by exchanging the two particles, leaving physical observables such as the probabilities $|\psi(\mathbf r_1,\mathbf r_2)|^2=|\psi(\mathbf r_2,\mathbf r_1)|^2$ of the two (indistinguishable) particles invariant. In summary, we see that the wavefunction can differ by a phase factor, 
$\psi(\mathbf r_2,\mathbf r_1)=e^{i\theta}\psi(\mathbf r_1,\mathbf r_2)$.

In three dimensions, there are only two possibilities for the phase since the operation of swapping twice (which means to move the first particle around the second one) can be continuously deformed to doing nothing, so we require $e^{i2\theta}=1$ leaving only the two possibilities $e^{i\theta}=1$ for \emph{bosons} and $e^{i\theta}=-1$ for \emph{fermions}. That is not true in two dimensions, so other phases are allowed, and \emph{any} phase is possible, which opens the existence of \emph{anyons} \cite{Lancaster2023}.
It turns out that electrons are fermions, and we will not consider the other cases further at this point.

Instead, we want to introduce a framework in which we do not need to describe the wavefunction of a many-particle state by (1) giving the position of all particles and (2) imposing the constraints that are needed to fulfill the exchange statistics. Instead, we use the second quantization to automatically enforce the statistics by using the occupation numbers of quantum states, i.e. stating how many particles are in a specific quantum state labeled by $\alpha_i$.
  The many-particle state is written as 
$|n_1,n_2,n_3,\ldots\rangle$
where $n_i$ is the number of particles that occupy the state $\alpha_i$. 
For fermions, the occupation numbers are $n_i \in \{0,1\}$ because having two or more such particles in the same state cannot acquire a minus sign by exchanging these two particles since the state with exchanged particles would be identical. This property is known as the  Pauli exclusion principle.

Instead of explicitly working  with many-particle states, one can introduce the creation  operators $c_\alpha^\dagger$ creating a fermion in state $\alpha$ and the annihilation operator $c_\alpha$ annihilating a fermion in state $\alpha$. The successive application of creation operators on the vacuum state $|0\rangle$ defines the many-particle state $|n_1,n_2,n_3,\ldots\rangle$. The key property of fermions that the sign of the wavefunction changes if exchanging two particles imposes the anticommutation relations of the fermionic operators
\begin{align}
\{c_\alpha,c_\beta^\dagger\}=\delta_{\alpha\beta}\quad 
\{c_\alpha,c_\beta\}=0\quad
\{c_\alpha^\dagger,c_\beta^\dagger\}&=0\label{eq_anticomm}
\end{align}
where the anticommutator of two operators is defined as $\{A,B\}=AB+BA$. 
This leads for example to $c_\alpha^\dagger c_\beta^\dagger=-c_\beta^\dagger c_\alpha^\dagger$, i.e. creating two fermions in opposite order which is equivalent to exchanging them, gives a minus sign to the state.
We mention at this point that the many-particle state that automatically fulfills Pauli statistics is given by the so-called Slater-determinant of ordinary many-particle states. For the following, this detail is not important; using the properties of the fermionic operators is sufficient.

Some special cases are obtained when setting $\alpha=\beta$ leading to $(c_\alpha)^2=0$ and  $(c_\alpha^\dagger)^2=0$, i.e. there is no state that has two fermions in the same state.

A Hamiltonian for free fermions can now be written for example as,
\begin{equation}
 H=\sum_{\alpha} \epsilon_{\alpha} c_\alpha^\dagger c_\alpha\label{eq_fermion}
\end{equation}
where $\epsilon_{\alpha}$ is a single particle energy.
Linear combinations of fermionic operators described by a unitary transformation are again fermionic operators and fulfill the anticommutation relations. With this, we can define fermion operators in momentum space $c_{\mathbf k,\alpha}=\frac{1}{\sqrt{N}} \sum_{i} c_{i,\alpha} e^{-i\mathbf k\cdot \mathbf R_i}$ and eventually fermions in band space where we use the eigenvector of the Bloch Hamiltonian to construct the corresponding linear combination
$c_{\mathbf k,n}=\sum_{\alpha} u^\alpha_n(\mathbf{k}) c_{\mathbf k,\alpha}$ where $u^\alpha_n(\mathbf{k})$ is a component of the eigenvector $u_n(\mathbf{k})$.

\subsection{Thermal expectation values}
\label{thermal_exp}
The expectation value as introduced in Eq.~(\ref{eq_pair_corr}) is taken in a many-particle quantum state.
Given an operator $O$, we obtain its expectation value via \begin{equation}
\langle O \rangle=\frac{\mathop\mathrm{Tr}(\exp(-\beta(H-\mu \hat N))O)}{\mathop\mathrm{Tr}(\exp(-\beta(H-\mu \hat N)))}
\end{equation}
Here $\beta=1/(k_\mathrm{B}T)$ is the inverse temperature, $H$ is the (many particle) Hamiltonian and $\mu$ is the chemical potential that fixes the number of electrons as expectation value of the particle number operator $\hat N=\sum_\alpha c^\dagger_\alpha c_\alpha$. The denominator is the partition function
$Z=\mathop\mathrm{Tr}(\exp(-\beta(H-\mu \hat N)))$.

As an example, we can calculate the occupation number of a single fermionic mode with the Hamiltonian $H=\epsilon c^\dagger c$. We obtain two allowed states $|0\rangle$ and $|1\rangle=c^\dagger |0\rangle$ and calculate the expectation value of the operator $n=c^\dagger c$.
The trace then contains exactly two terms that yield the partition function
$ Z=\sum_{n=0}^1 \langle n |\exp(-\beta(H-\mu \hat N))|n \rangle=1+\exp(-\beta (\epsilon -\mu))
$ 
and for the numerator just one term contributes, such that the final result is the Fermi function, 
\begin{equation}\langle n\rangle=\frac{1}{e^{\beta(\epsilon -\mu)} +1}=f(\epsilon)
\end{equation}
For a quadratic fermionic Hamiltonian as in Eq.~(\ref{eq_fermion}), one can calculate the thermal averages with the result
\begin{equation}
\langle c_\alpha^\dagger c_\beta\rangle = \delta_{\alpha,\beta} f(\epsilon_\alpha)\label{thermal_expectation}
\end{equation}
and by the use of the anticommutation relation
$\langle c_\alpha c_\beta^\dagger\rangle = \delta_{\alpha,\beta} (1-f(\epsilon_\alpha)).$
We conclude the discussion by observing that such expectation values inherit the properties of the operators that they are calculated from, i.e. symmetries on quantum numbers are preserved even if the expectation value is taken in a thermal quantum state.

\section{Solving the linearized gap equation}
\label{sec_lge_details}
Starting from the pairing interaction given in Eq.~(\ref{eq_tJ}), we obtain the singlet and triplet interactions from the (anti-) symmetrization, Eq.~(\ref{eq_pairing_symm}), $V^{s}(\kv,\kv')=U-\frac{3J}{2}(\cos k_x \cos k_x'+\cos k_y \cos k_y')$ and 
$V^{t}(\kv,\kv')=-\frac{3J}{2}(\sin k_x \sin k_x'+\sin k_y \sin k_y')$.
The order parameters in the point group $D_4$ can be expanded using the functions $\phi_{A_1}^0= 1$, $\phi_{A_1}^1(\kv)=\cos k_x+\cos k_y$, $\phi_{B_1}(\kv)= \cos k_x-\cos k_y$, $\phi_{E_x}(\kv)=\sqrt{2}\sin k_x$,  $\phi_{E_y}(\kv)=\sqrt{2}\sin k_y$ where we note that we need two functions for $A_1$ and for the given pairing interaction, there is no contribution to $A_2$ and $B_2$.

Next, we rewrite $V^{s}(\kv,\kv')=V_{A_1}+V_{B_1}$ with
the $A_1$ term consisting in a repulsive part and the extended $s$-wave interaction
$V_{A_1}=U-\frac{3J}{4}\phi_{A_1}^1(\kv)\phi_{A_1}^1(\kv')$ and the $B_1$ interaction
$V_{B_1}=-\frac{3J}{4}\phi_{B_1}(\kv)\phi_{B_1}(\kv')$. The antisymmetrized triplet interaction 
$V^{t}=V_{E_x}+V_{E_y}$ can be decomposed into the contributions of the two basis functions of the $E$ irrep with
$V_{E_i}=-\frac{3J}{4}\phi_{E_i}(\kv)\phi_{E_i}(\kv')$.

Inserting in Eq.~(\ref{eq_self}) and linearizing the self-consistency equation yields for the $B_1$, $E_x$ and $E_y$ irreps
\begin{equation}
 1=\frac{3J}{4}\sum_\kv \frac{\phi_i(\kv)^2}{2\epsilon_\kv}\tanh\Big(\frac{\beta \epsilon_{\kv}}{2}\Big).
\end{equation}
We note at this point that the function $\chi(\epsilon)=\frac {\tanh(\beta \epsilon/2)}{2\epsilon}$ indeed has its maximum at $\epsilon=0$, i.e. there are significant contributions from the states close to the Fermi level, but the function only reaches a constant $\lim_{\epsilon\rightarrow 0}\chi(\epsilon)= \beta/4$ increasing with lower temperature, but also has a tail that only falls off with $\chi(\epsilon) \sim 1/(2|\epsilon|)$ at large $|\epsilon|$. For the $A_1$ channel, the situation is a bit more complicated because there are two possible basis functions, such that we need to factor out the product $V_{A_1} (\Delta_{A_1}^0 +\Delta_{A_1}^1\phi_{A_1}^1(\kv))$ under the integral. Defining the moments $I_n=\sum_\kv \frac{\phi^1_{A_1}(\kv)^n}{2\epsilon_\kv}\tanh\Big(\frac{\beta \epsilon_{\kv}}{2}\Big)$, $n=0,1,2$, the equation becomes a matrix equation
\begin{equation} 
 \left(\begin{array}{c}
  \Delta_{A_1}^0\\
    \Delta_{A_1}^1
 \end{array}\right)
= \left(\begin{array}{cc}
  -U I_0& -U I_1\\
  \frac{3J}{4}I_1 & \frac{3J}{4}I_2
 \end{array}\right)
 \left(\begin{array}{c}
  \Delta_{A_1}^0\\
    \Delta_{A_1}^1
 \end{array}\right)
\end{equation}
with the solution that the matrix has the (largest) eigenvalue $1$ at $T_c$. Indeed, for $J=0$ and $U<0$ we recover the case as discussed in section~\ref{sec_conventional}, while for $U>0$ which describes the remaining part of the Coulomb interaction, it depends whether there is a solution to this equation. If the first moment vanishes, $I_1=0$ , i.e. $\phi^1_{A_1}(\kv)$ averages to zero when weighted accordingly, the solution just exhibits this component, while for finite $I_1\neq 0$, the system chooses to have both components to avoid the repulsive part of the Coulomb interaction.

\end{document}